\documentclass{siamart171218}
\usepackage{braket}
\usepackage{amsmath,amssymb,amsfonts}
\usepackage{algorithmic}
\usepackage{graphicx}
\usepackage{textcomp}

\usepackage{color} 
 \usepackage{amsfonts}
 \usepackage{amsmath}
 \usepackage{amssymb}
 \usepackage{array}
 \usepackage{mathrsfs}
 \usepackage{leftidx}
 \usepackage{graphicx}
 \usepackage{epstopdf}
 \usepackage{dcolumn}
 \usepackage{bm}

\usepackage{cite}

\usepackage{multirow}

\usepackage{cases}

\usepackage{lipsum}
\usepackage{amsfonts}
\usepackage{graphicx}
\usepackage{epstopdf}
\usepackage{algorithmic}
\ifpdf
  \DeclareGraphicsExtensions{.eps,.pdf,.png,.jpg}
\else
  \DeclareGraphicsExtensions{.eps}
\fi

\newsiamremark{remark}{Remark}
\newsiamremark{hypothesis}{Hypothesis}
\crefname{hypothesis}{Hypothesis}{Hypotheses}
\newsiamthm{claim}{Claim}

\def\BibTeX{{\rm B\kern-.05em{\sc i\kern-.025em b}\kern-.08em
    T\kern-.1667em\lower.7ex\hbox{E}\kern-.125emX}}

\title{Quantum coherent and measurement feedback control based on atoms coupled with a semi-infinite waveguide\thanks{\funding{This work is partially supported by the ANR project  Q-COAST ANR-19-CE48-0003, the ANR project IGNITION ANR-21-CE47-0015, Innovation Program for Quantum Science and Technology 2023ZD0300600, Guangdong Provincial Quantum Science Strategic Initiative (No. GDZX2200001), Hong Kong Research Grant Council (RGC) under Grant No. 15213924, National Natural Science Foundation of China under Grants Nos.  62173288.}}}
\author{Haijin Ding\thanks{Corresponding author (H. Ding). Laboratoire des Signaux et Syst\`{e}mes (L2S), CNRS-CentraleSup\'{e}lec-Universit\'{e} Paris-Sud, Universit\'{e} Paris-Saclay, 3, Rue Joliot Curie, 91190, Gif-sur-Yvette, France, and Department of Applied Mathematics, The Hong Kong Polytechnic University, Kowloon 999077, Hong Kong, China. (\email{dhj17@tsinghua.org.cn}).} \and Nina H. Amini\thanks{Laboratoire des Signaux et Syst\`{e}mes (L2S), CNRS-CentraleSup\'{e}lec-Universit\'{e} Paris-Sud, Universit\'{e} Paris-Saclay, 3, Rue Joliot Curie, 91190, Gif-sur-Yvette, France. 
  (\email{ nina.amini@l2s.centralesupelec.fr}).} \and Guofeng Zhang\thanks{Corresponding author (G. Zhang). Department of Applied Mathematics, The Hong Kong Polytechnic University, Kowloon 999077, Hong Kong, China, and The Hong Kong Polytechnic University Shenzhen Research Institute, Shenzhen, 518057, China (\email{guofeng.zhang@polyu.edu.hk}).}
\and John E. Gough\thanks{Aberystwyth University, SY23 3BZ, Wales, United Kingdom (\email{ jug@aber.ac.uk}).}}

\headers{Quantum coherent and measurement feedback control}{Haijin Ding, Nina H. Amini, Guofeng Zhang and John E. Gough}

\ifpdf
\hypersetup{
  pdftitle={An Example Article},
  pdfauthor={H. J. DING, and }
}
\fi

\usepackage[a4paper,centering]{geometry}

\begin{document}

\maketitle


\begin{abstract} In this paper, we demonstrate the application of quantum feedback control in creating desired states for atomic and photonic systems utilizing a semi-infinite waveguide coupled with multiple two-level atoms. In this approach, an initially excited atom can emit one photon into the waveguide, and the photon can be either reflected by the terminal mirror of the waveguide or other atoms, establishing various feedback loops through the coherent interactions between the atom and photon. When there are no more than two excitations in the waveguide quantum electrodynamics (QED) system, the evolution of quantum states can be effectively elucidated through the lens of random graph theory. However, this process is influenced by the environment. We propose that the environment-induced dynamics in the coherent feedback loop can be eliminated by measurement-based feedback control or coherent drives. Consequently, in the atom-waveguide interactions in open quantum systems, measurement-based feedback has the potential to modulate the quantum steady states. Additionally, the homodyne detection noise during the measurement process can induce errors upon quantum states, which can be influenced by the coherent feedback parameter designs.
\end{abstract}

\begin{keywords}
quantum feedback control, coherent feedback, measurement feedback, atom-waveguide interaction
\end{keywords}

\begin{AMS}
81Q93, 81S25, 93B52, 93C43
\end{AMS}

\section{Introduction}
Quantum feedback control has attracted much attention due to its wide applications in quantum information processing (QIP)~\cite{zhang2017quantum,LloydCoherent}, quantum optics~\cite{feedbackQOreview,zhang2017quantum,zhang2012quantum}, and quantum metrology~\cite{fallani2022learning,al2022quantum,salvia2022critical}. In contrast to open-loop quantum control where the quantum state evolves unitarily regulated by control pulses pre-designed based on optimal control theory~\cite{itami2005nonlinear,altafini2012modeling,d2024optimal,james2004risk,nurdin2009coherent,ho2010accelerated,wu2018data} or robust optimization~\cite{wang2023quantum,koswara2021robust,weidner2023robust,wu2019learning,ding2019robust,ding2021collaborative,dong2022dynamics,ZD22}, quantum feedback control offers an alternative calibration approach with high efficiency; see e.g., \cite{bardeen1997feedback,assion1998control,wiseman2009quantum,altafini2012modeling,ZLH+12,zhang2017quantum}.
Depending on whether or not the measurement results of the underlying quantum system are utilized in the feedback control laws, quantum feedback can be classified into measurement-based and coherent quantum feedback. In the coherent feedback design, the quantum plant can be directly coupled with another quantum component which acts as a controller~\cite{LloydCoherent,ZJ11,ZJ12}, therefore leading to a unitary evolution of the plant-controller system. 
For instance,  the nonlinear crystal in an optical cavity can be pumped by an input field, while the output field can be made to re-interact with the crystal by reflecting it back via the mirrors forming the cavity, thus effectively setting up a quantum coherent feedback loop~\cite{PengKCCFC}. The coherent feedback mechanism avoids introducing backaction noise into the quantum system and allows for adjustable feedback loop length, making it an excellent mechanism for generating optical entangled states. In contrast, measurement-based quantum feedback control relies on the measurement results of the quantum system to manage the nonlinear conditioned evolution of the quantum state. The measurement process can introduce noises into the evolution of quantum states, distinguishing it from the open-loop control methods~\cite{stefanatos2021shortcut,koch2022quantum,wu2019learning,ding2019robust,ding2021collaborative,dong2022dynamics,wu2018data} and coherent feedback. 
For example, in quantum error correction by means of encoded three-qubit bit-flip codes, quantum states with evolution errors can be estimated via continuous measurement, based on which appropriate feedback action can be applied to each qubit to correct the operation errors~\cite{MFQECPractical}.

Quantum coherent feedback control involves interactions among the atom, cavity, waveguide, etc. For example, two cavities may be coupled via a semi-transparent mirror, allowing an excited atom in one of the cavities to emit a coherent field (i.e., a photon) into the other cavity. Subsequently, the photon can re-interact with the atom in the former cavity, thereby constructing a quantum coherent feedback loop~\cite{ScullyTreatment,spacetime,CavityEquation}.
Alternatively, a cavity containing one or more atoms can also be coupled with a semi-infinite waveguide of continuous modes. Photons in the cavity can be emitted into the waveguide and re-interact with cavity-QED system after propagating along the waveguide~\cite{photonfeedback}. Thus, the dynamics of the cavity-QED system (i.e., the atomic states and the number of photons) can be influenced by the parameters of the coherent feedback loop such as the length of the waveguide and cavity-waveguide coupling strength. In another scenario, an atom can be directly coupled with the waveguide without the mediation of the cavity~\cite{DingZhangAutomatica}. In this case, the emission and reflection of photon form  a coherent feedback loop, and the dynamics  is affected by the coupling strength between the atom and the waveguide and the transmission delay of photons in the waveguide. When there are multiple atoms coupled with a waveguide, a photon emitted by one atom can be absorbed and re-emitted by another after being transmitted in the waveguide, thus forming multiple coherent feedback loops among atoms~\cite{twoatomtwophoton}. The complexity of coherent feedback control increases if the system has many  emitters and excitations, or couplings between the atom and the waveguide are chiral~\cite{Zoller2016PRL,bound1,qiao2019atom,DingZhangAutomatica} . Chiral couplings can dictate the propagation direction of the emitted photonic wave packets in the waveguide~\cite{DingZhangAutomatica}.  For a system where two two-level atoms are coupled with a semi-infinite waveguide but the whole system has only one excitation, coherent feedback dynamics can induce entanglement between the two atoms~\cite{ZhangBin}. In a coherent feedback network with two excitations, the creation of two-photon states depends on the positioning of the two atoms in the a semi-infinite waveguide~\cite{DingZhangAutomatica}. When there are multiple atoms coupled with an infinite waveguide, the entanglement between arbitrary two atoms can be modulated by tuning the locations of atoms and their chiral couplings with the waveguide~\cite{pichler2015quantum}.

In quantum measurement feedback control, the information obtained from measuring the quantum system can be utilized to regulate the quantum system~\cite{zhang2017quantum}. For instance, in a simple one-bit toy model for quantum computation~\cite{MFQECPractical}, the random bit-flip error can alter the encoded qubit's state. Subsequently, using the measurement information of the quantum system, the feedback control can steer the qubit towards the desired state.  Throughout this process, quantum measurements can introduce random noise and the quantum system dynamics is described by a stochastic master equation~\cite{wiseman1994quantum,GZP19,GZP20}. The closed-loop  feedback control performance of the qubit depends on the measurement and feedback strengths, the detector efficiency, etc., and can be optimized by tuning a smoothing filter for measurement results~\cite{MFQECPractical,borah2022measurement}. A commonly used approach is the three-qubit bit-flip coding for quantum error correction~\cite{MFQECPractical,QECBeginner,threeBitQECCavity}, where the measurement feedback control method generalized from the one-qubit case can detect and correct not only a single bit-flip error, but also the double flipping error. Quantum measurement feedback control is also helpful for the generation of entanglement between two qubits, with a higher fidelity compared with open-loop control~\cite{martin2017optimal,sarovar2005high}. As proposed in~\cite{HamFeedEntangle}, a feedback Hamiltonian designed based on weak measurement can steer a quantum system to a decoherence-free subspace, and this can further enhance the bipartite entanglement. 
Moreover, in an open quantum system suffering from relaxation and dephasing effects, properly designed measurement feedback can preserve the quantum system's coherence~\cite{zhang2010protectingTAC}.

Quantum measurement can be categorized as quantum non-demolition
(QND) measurement and non-QND measurement, depending on whether the measurement operator commutes with the quantum system Hamiltonian or not\cite{QNDgrangier1998quantum,qi2012TAC,xiaQND,RouchonAutomatica,amini2023exponential,liang2019exponential}. For instance, in the early implementation of photon QND measurement using the cavity-QED system, an atom serving as a meter can interact with the cavity by absorbing and emitting a photon when there is initially one photon in the cavity. Subsequently, the existence of photons in the cavity can be deduced by observing the phase shift of the atom wavefunction~\cite{nogues1999seeing}. In this process, the atom meter does not annihilate the photon in the cavity, and it can further be used to stabilize the photon number in the cavity~\cite{Nature2011Real}. However, when the measurement operator does not commute with the system Hamiltonian, theoretically some eigenstates of the system Hamiltonian cannot be generated no matter how the control fields are designed~\cite{qi2012TAC}. Non-QND measurement is achievable within the waveguide QED framework, where atoms can be coupled with an infinite or semi-infinite waveguide~\cite{OPSB,zhan2022long,buonaiuto2021dynamical,gonzalez2017efficient,olmos2020interaction,strandberg2019numerical,Zoller1}, and homodyne detection or photon counting at the output end of the waveguide can be implemented. Similar to the cavity-QED system, the homodyne detection results can be used to design the feedback control fields for the waveguide QED system.

The dynamics of an atom coupled with a semi-infinite waveguide has been studied in~\cite{bradford2013spontaneous,tufarelli2014non,barkemeyer2021strongly,bound1,dorner2002laser}, where the spontaneous emission of the atom is inhibited if the emitted wavelength matches the atom-mirror distance, thus creating an atom-photon bound state. A two-atom network is studied in~\cite{DingZhangAutomatica,ZhangBin}, where it is shown that the coherent feedback and chiral couplings between the atoms and waveguide can influence the generation of single-photon and two-photon states and the entanglement between the two atoms.
In Section \ref{sec:coherentgraph} of this paper, we study the coherent feedback dynamics of a multi-atom network chirally coupled with a semi-infinite waveguide, and propose that this multi-atom coherent feedback network containing one or two photons can be mapped to a multi-agent network on a random graph. The consensus of classical multi-agent networks has been studied in~\cite{ConsensusDefTAC,consensusTAC}. It was introduced in~\cite{mazzarella2014consensus} that quantum consensus can be realized in a multi-qubit network if the populations of different qubits are the same. In this paper, we show that the above quantum consensus can be realized with the multi-atom network coupled with the semi-infinite waveguide by tuning the atoms' locations and their non-chiral couplings with the waveguide.
Moreover, in Section \ref{sec:hybridopen}, we study the quantum coherent feedback dynamics influenced by the environment, generalizing the results established for closed quantum systems in~\cite{DingZhangAutomatica,bound1,qiao2019atom} to open quantum systems. As studied in~\cite{bradford2013spontaneous,WZH2022}, the above results for closed quantum systems do not hold if there are no drives applied to the atoms. Therefore, in this paper, we explore the possibility of exciting the atom coupled with a semi-infinite waveguide using coherent drives or quantum measurement feedback. Additionally, in the quantum measurement feedback for atom-waveguide interactions introduced in~\cite{strandberg2019numerical,zhan2022chirality,guimond2016chiral,OPSB}, the control field applied to the atom can be designed using homodyne detection. Based on this, we further illustrate how the stochastic atomic dynamics under measurement feedback can be affected by the chiral couplings between the atom and waveguide.
We summarize the main contributions as follows.
\begin{itemize}
\item In \textbf{Theorem~\ref{consensus}}, we propose a condition for reaching quantum consensus when multiple atoms are coupled with a semi-infinite waveguide.
\item In \textbf{Theorem~\ref{Theorem3Finalstate}} and \textbf{Theorem~\ref{Theorem5Y0}}, we show how the steady atomic state in the open system can be influenced by an external drive and a coherent feedback loop closed by the semi-infinite waveguide.
\item In \textbf{Theorem~\ref{Theorem6MF}}, we show how a measurement feedback control can affect the steady quantum states.\
\item In \textbf{Theorem~\ref{Cohestability}} and \textbf{Theorem~\ref{transition}}, we clarify how a coherent feedback design can affect the quantum state's robustness to the measurement noise and the convergence rate in the transient process. 
\end{itemize}

The paper is organized as follows. In Section~\ref{sec:coherentgraph}, we investigate the coherent feedback control of the quantum network as shown in Fig.~\ref{fig:NatomWaveguide} (a). In particular, we show that the evolution of the quantum states can be described using ideas from random graph theory. In Section~\ref{sec:hybridopen} we apply both the coherent feedback control and measurement feedback control to investigate the dynamics of the  the quantum network in Fig.~\ref{fig:NatomWaveguide} (a). We conclude the paper in Section~\ref{sec:conclusions}.

\section{Coherent feedback control with one or two excitations}\label{sec:coherentgraph}

\begin{figure}[htbp]
  \centering
  \centerline{\includegraphics[width=1\columnwidth]{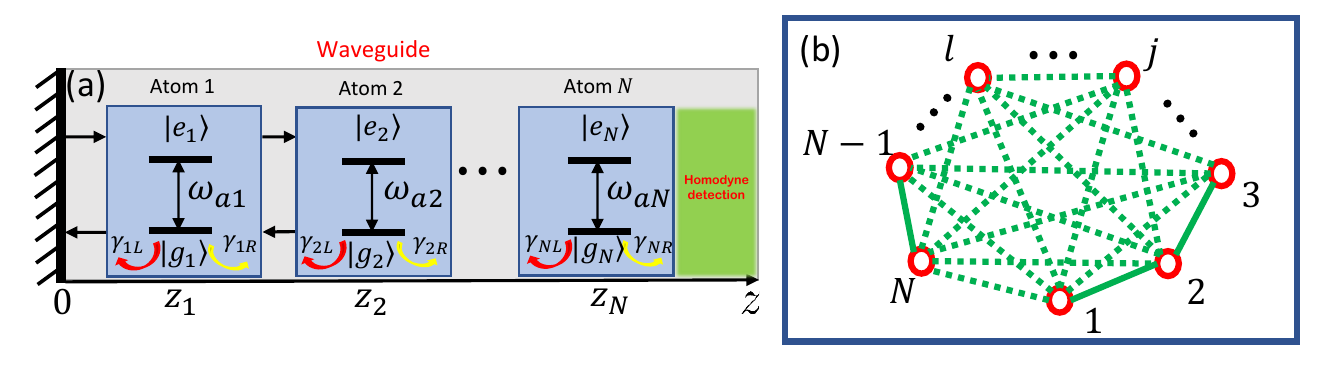}}
  \caption{(a) Feedback control of $N$ two-level atoms coupled with a semi-infinite waveguide. (b) Random graph with $N$ nodes.}
  \label{fig:NatomWaveguide}
\end{figure}

As shown  in Fig.~\ref{fig:NatomWaveguide} (a),  $N$ two-level atoms are coupled to a semi-infinite waveguide with a perfect reflecting mirror at $z=0$, and the distance between the mirror and the $j$th atom of the resonant frequency $\omega_{aj}$ is $z_j$. The coupling strengths between the $j$th atom and the right/left  propagating fields in the waveguide are $\gamma_{jR}$ and $\gamma_{jL}$, respectively. The interaction Hamiltonian between the $j$th atom and the waveguide reads 
\begin{equation} \label{con:Hintj2nonchiral}
\begin{aligned}
H_{I}^{(j)} 
&= \int  \left [g_{kjt}(k,t,z) d^{\dag}_k\sigma_j^-  + g_{kjt}^*(k,t,z) d_k\sigma_j^+\right ]\mathrm{d}k,
\end{aligned}
\end{equation}
where $d_k$($d^{\dag}_k$) are the annihilation (creation)
operators of the propagating waveguide mode $k$, $\sigma_j^- = |g_j\rangle \langle e_j|$ and $\sigma_j^+ = |e_j\rangle \langle g_j|$ are the lowering and raising operators of the $j$th atom, the coupling strength $g_{kjt}$ between the $j$th atom and the waveguide is
\begin{equation}\label{eq:g_kjt}
\begin{aligned}
g_{kjt}(k,t,z)&= i\left (\gamma_{jR} e^{-i\omega_k z_j/c}-\gamma_{jL} e^{i\omega_k z_j/c}\right) e^{i(\omega_k-\omega_{aj})t} \\
&= i\gamma_{jR} e^{i\left[(\omega_k-\omega_{aj})t - \omega_k z_j/c\right ]} - i\gamma_{jL}e^{i\left [(\omega_k-\omega_{aj})t + \omega_k z_j/c\right ]},
\end{aligned}
\end{equation}
where $\omega_k = kc$ with $c$ being the speed of light in the waveguide.

When the coupling between the atoms and the waveguide is nonchiral, that is $\gamma_{jR} = \gamma_{jL} \equiv \gamma_{j}$, the coupling in Eq. \eqref{eq:g_kjt} reduces to $g_{kjt} = 2\gamma_j \sin(kz_j) e^{i(\omega_k-\omega_{aj})t}$. The output field  at the right end of the waveguide can be measured via the homodyne detection  ~\cite{OPSB,strandberg2019numerical,PRLHOmNatom}. However, in this section, we ignore the measurement process. 
We will explore how the measurement feedback can control the atomic dynamics in Section \ref{sec:hybridopen}.

In this section, we investigate the quantum coherent feedback control dynamics in the interaction picture with the Hamiltonian $H =\sum_{j=1}^N H_I^{(j)} $. The following two assumptions are used.

\newtheorem{assumption}{Assumption}
\begin{assumption} \label{AsumInitialB}
In this section, initially the waveguide is empty, and there is one or two atoms in the excited state.
\end{assumption}

\begin{assumption} \label{AsumInteraction}
The atoms have the same resonant frequency   $\omega_{a1} = \omega_{a2}=\cdots=\omega_{aN} \equiv  \omega_{a}$. There exist no direct couplings among the atoms.  
\end{assumption}

Based on \textbf{Assumption}~\ref{AsumInitialB} and  \textbf{Assumption}~\ref{AsumInteraction}, the populations of eigenstates of the quantum system in Fig.~\ref{fig:NatomWaveguide} (a) can be represented in terms of the evolution of the random graph in Fig.~\ref{fig:NatomWaveguide} (b). For example, when there is only one excitation in this  quantum network, the probability of the $j$th atom being excited can be interpreted as the probability of the $j$th vertex of the random graph. When there are  two excitations, the probability of two atoms (e.g., the $j$th and $l$th atoms) being simultaneously excited can be interpreted as the probability of the edge connecting the $j$th and $l$th vertices in the random graph, while the probability that only the $j$th atom is excited and there is one photon in the waveguide can be interpreted as the probability of the $j$th vertex. 

\subsection{Coherent feedback control with one excitation}

When there is a single excitation in quantum system in Fig.~\ref{fig:NatomWaveguide} (a), the quantum state can be represented as
\begin{equation} \label{con:State}
\begin{aligned}
|\Psi(t)\rangle =& \sum_{j=1}^N c_j(t)|g_1,g_2,\cdots,g_{j-1},e_{j},g_{j+1},\cdots, g_{N},\{0\}\rangle \\
& + \int \tilde{c}(t,k) |g_1,\cdots, g_{N},\{k\}\rangle \mathrm{d}k, \\
\end{aligned}
\end{equation}
where the first component  on the right-hand side (RHS) describes the state where one atom (the $j$th, for $j=1,2,\cdots,N$) is excited while there are no photons in the waveguide, and the second component represents that all the atoms are at the ground state and there is one photon in the waveguide. For ease of discussion, in what follows
we assume that initially only the {\it first} atom is excited.

\begin{remark}
For  the quantum network in Fig.~\ref{fig:NatomWaveguide} (a) with one excitation, the $j$th vertex on the random graph in Fig.~\ref{fig:NatomWaveguide} (b) with probability $p_j$ represents the probability that the $j$th two-level atom is excited, namely $p_j = |c_j|^2$ with $\sum_j^N p_j \leq1 $.
\end{remark}

The dynamics of the quantum state is governed by the Schr\"{o}dinger equation
\[
\frac{\mathrm{d}}{\mathrm{d} t}|\Psi(t)\rangle = -i H |\Psi(t)\rangle.
\]
Based on \textbf{Assumption}~\ref{AsumInitialB}, the evolution of the amplitudes in Eq. \eqref{con:State} is given by
\begin{small}
\begin{subequations} \label{con:WavNatom}
\begin{numcases}{}
\dot{c}_j(t)
= \int \left \{ -\gamma_{jR} e^{-i[(\omega_k-\omega_a)t - \omega_k z_j/c]} +\gamma_{jL}e^{-i[(\omega_k-\omega_a)t + \omega_k z_j/c]}\right\}\tilde{c}(t,k) \mathrm{d}k,   \label{cjt}\\
\dot{\tilde{c}}(t,k)
= \sum_{j=1}^N  c_j(t) \left \{ \gamma_{jR} e^{i[(\omega_k-\omega_a)t - \omega_k z_j/c]} - \gamma_{jL}e^{i[(\omega_k-\omega_a)t + \omega_k z_j/c]}\right\}. \label{ckt}
\end{numcases}
\end{subequations}
\end{small}%
 Eq.~(\ref{cjt}) means that the $j$th atom  can absorb one photon from the waveguide and thus is excited. Eq.~(\ref{ckt}) means that each excited two-level atom can emit one photon into the waveguide via the spontaneous emission. Clearly, all the amplitudes are influenced by the locations of atoms and the chiral coupling strengths between atoms and the waveguide. According to the calculations given in Appendix~\ref{Sec:ApendixDelayOnexcitation}, Eq.~(\ref{con:WavNatom}) can be rewritten as a delay-dependent ODE 
\begin{small}
\begin{equation} \label{con:WavNatomDelay}
\begin{aligned}
&\dot{c}_j(t) = - \frac{\gamma_{jR}^2 + \gamma_{jL}^2 }{2}c_j(t) -\gamma_{jR} \sum_{p=1}^{j-1}  \gamma_{pR}  c_p\left (t-\frac{z_j-z_p}{c}\right )e^{i\omega_a\frac{z_j-z_p}{c}} \\
&-\gamma_{jL} \sum_{p=j+1}^N \gamma_{pL} c_p\left (t-\frac{z_p-z_j}{c}\right )  e^{i\omega_a\frac{z_p-z_j}{c}} +\gamma_{jR} \sum_{p=1}^N  \gamma_{pL} c_p\left (t-\frac{z_p+z_j}{c}\right )  e^{i\omega_a \frac{z_p+z_j}{c}} .  
\end{aligned}
\end{equation}
\end{small}%
The first component of RHS of Eq.~(\ref{con:WavNatomDelay}) means that the $j$th atom can decay to its ground state by emitting a photon along the right and left directions. The second component represents that the excited state can be transferred from the $p$th atom to the $j$th atom along the right direction with a delay determined by the distance between two atoms; similarly for the left direction in the third component. The fourth component represents the transmission of the excited state after the reflection by the mirror with a round trip delay.

\subsubsection{Feedback network with single delay}

For the simplified circumstance that $z_1 = z_2 = \cdots = z_N$, i.e., all the atoms are in the same place, denote 
\[
x(t) = \left [c_1(t), c_2(t), \cdots, c_N(t) \right ]^{\mathrm{T}}
\] 
with the superscript $\mathrm{T}$ stands for matrix transpose.  Then Eq.~(\ref{con:WavNatomDelay}) reduces to
\begin{equation} \label{con:OneExciationVector}
\begin{aligned}
\dot{x}(t) &= \begin{bmatrix}
  - \frac{\gamma_{1R}^2 + \gamma_{1L}^2 }{2} &-\gamma_{1L}\gamma_{2L} &-\gamma_{1L}\gamma_{3L} &\cdots &-\gamma_{1L}\gamma_{NL}\\
  -\gamma_{1R}\gamma_{2R} &- \frac{\gamma_{2R}^2 + \gamma_{2L}^2 }{2} &-\gamma_{2L}\gamma_{3L} &\cdots  &-\gamma_{2L}\gamma_{NL}\\
   \vdots &\vdots &\vdots &\ddots &\vdots\\
  -\gamma_{1R}\gamma_{NR} &-\gamma_{2R}\gamma_{NR} &-\gamma_{3R}\gamma_{NR}  & \cdots  & - \frac{\gamma_{NR}^2 + \gamma_{NL}^2 }{2}
  \end{bmatrix}  x(t)\\
  & ~~~~+ e^{i\omega_a \tau} \begin{bmatrix}
\gamma_{1R}\\ \gamma_{2R} \\ \vdots \\ \gamma_{NR}
  \end{bmatrix}
  \begin{bmatrix}
\gamma_{1L}  & \gamma_{2L} & \cdots & \gamma_{NL}
  \end{bmatrix}
  x(t-\tau)\Theta(t-\tau)\\
  &\triangleq A_0x(t) +  e^{i\omega_a \tau} B_0 x(t-\tau)\Theta(t-\tau),
\end{aligned}
\end{equation}
where $\tau = 2z_1/c$ represents the round trip delay and $\Theta(t)$ is the Heaviside step function. Denote $\Gamma_R = \left [\gamma_{1R}, \gamma_{1R}, \cdots, \gamma_{NR} \right ]^T$ and $\Gamma_L = \left [\gamma_{1L}, \gamma_{1L}, \cdots, \gamma_{NL} \right ]^T$. Then the matrix  $B_0 = \Gamma_R \otimes \Gamma_L^T$ with $\otimes$ representing the Kronecker product.

The Laplace transform of a time-domain signal  $x(t)$ is denoted as $X(s)$ in the frequency domain. Applying the Laplace transform  to Eq. \eqref{con:OneExciationVector} yields
\begin{equation} \label{con:OneExciationLaplace}
\begin{aligned}
sX(s)-x(0) = A_0X(s) +  e^{i\omega_a \tau} B_0 e^{-s\tau} X(s),
\end{aligned}
\end{equation}
 which can be re-written as
\begin{equation} \label{con:XsSolution}
\begin{aligned}
X(s) = \left (sI - A_0 - e^{i\omega_a \tau} B_0 e^{-s\tau} \right  )^{-1}x(0).
\end{aligned}
\end{equation}

Denote  the Laplace transform of $c_j(t)$ by $C_j(s)$ and let
\[
\mathcal{C}_j =\left [\underbrace{  0,\cdots,0}_{j-1 ~\text{columns}}, 1,\underbrace{  0,\cdots,0}_{N-j ~\text{columns}}\right ].
\] 
Then from Eq. \eqref{con:XsSolution} we have
\begin{equation} \label{con:CjsSolution}
\begin{aligned}
C_j(s) =\mathcal{C}_j \left (sI - A_0 - e^{i\omega_a \tau} B_0 e^{-s\tau} \right  )^{-1}x(0).
\end{aligned}
\end{equation}
As only the first atom is  initially in the excited state, $x(0) = [1,0,\cdots,0]^T$. Denote $\mathcal{M} = sI - A_0 - e^{i\omega_a \tau} B_0 e^{-s\tau}$. Then $C_j(s) =\frac{1}{|\mathcal{M} |}M^*_{j1}$ with $M^*_{j1}$ being the cofactor  and $|\mathcal{M} |$ the determinant of the matrix $\mathcal{M}$.

\newtheorem{myDef}{Definition}
\begin{myDef} \label{ConsencusOne}
The quantum network with the  vertex set $\mathcal{V}$ as shown in Fig.~\ref{fig:NatomWaveguide} (b) reaches amplitude consensus at $t$ when~\cite{ConsensusDefTAC,consensusTAC}
\[
| c_j(t) - c_p(t) | = 0,
\]
for all $j,p \in \mathcal{V}$. The network reach population consensus at $t$ when~\cite{ConsensusDefTAC,consensusTAC}
\[
 |c_j(t)|^2 - |c_p(t)|^2  = 0,
\]
for all $j,p \in \mathcal{V}$.
\end{myDef}

Denote 
\[
\tilde{x}(t) = \left [\textbf{R}\left [ c_1(t) \right],  \textbf{I}\left [ c_1(t) \right], \textbf{R}\left [ c_2(t) \right],  \textbf{I}\left [ c_2(t) \right]\cdots, \textbf{R}\left [ c_N(t) \right], \textbf{I} \left [ c_N(t) \right]\right ]^{\mathrm{T}},
\] 
where $\textbf{R}$ and $\textbf{I}$ represent the real and imaginary parts of a complex number, respectively. Denote $\tilde{x}_j(t) = \left [\textbf{R}\left [ c_j(t) \right],  \textbf{I}\left [ c_j(t) \right]\right ]$, then $\tilde{x}(t) = \left [ \tilde{x}_1(t), \tilde{x}_2(t), \cdots, \tilde{x}_N(t)\right ]^{\mathrm{T}}$.
Then Eq. \eqref{con:OneExciationVector} can be rewritten as
\begin{equation} \label{con:TensorRealImag}
\begin{aligned}
\dot{\tilde{x}}(t)  =\left( A_0 \otimes I_2\right )\tilde{x}(t) + \left (  B_0 \otimes G(\tau) \right ) \tilde{x}(t-\tau),
\end{aligned}
\end{equation}
where $G(\tau) = \begin{bmatrix} -\cos\left(\omega_a \tau\right ) & \sin\left(\omega_a \tau\right ) \\ -\sin\left(\omega_a \tau\right ) & -\cos\left(\omega_a \tau\right )\end{bmatrix}$. 

Performing the Laplace transform on  Eq.~(\ref{con:TensorRealImag}) yields
\begin{equation} \label{con:TensorRealImagLaplace}
\begin{aligned}
s\tilde{\mathbf{x}}(s) - \tilde{x}(0)  =\left( A_0 \otimes I_2\right )\tilde{\mathbf{x}}(s) + \left ( B_0 \otimes G(\tau)\right ) e^{-s\tau} \tilde{\mathbf{x}}(s),
\end{aligned}
\end{equation}
which alternatively is
\begin{equation} \label{con:TensorRealImagLaplace2}
\begin{aligned}
\tilde{\mathbf{x}}(s) &=\left [ s\left( I_N   \otimes I_2\right ) - \left( A_0 \otimes I_2 \right ) -  \left ( B_0 \otimes  G(\tau) \right ) e^{-s\tau} \right ]^{-1}\tilde{x}(0)\triangleq \Delta_s ^{-1}\tilde{x}(0).
\end{aligned}
\end{equation}
\begin{remark}
In particular, when $\omega_a \tau = n\pi$ with $n=1,2,\cdots$, Eq. \eqref{con:TensorRealImagLaplace2} becomes $\tilde{\mathbf{x}}(s) =\left [ \left( s I_N  -A_0 \pm B_0 e^{-s\tau}\right )  \otimes I_2 \right ]^{-1}\tilde{x}(0)$, which means that the real and imaginary parts of $x(t)$ can be written in the format of tensors.
\end{remark}

\newtheorem{thm}{Theorem}
\begin{thm}\label{consensus}
When $\omega_a \tau =2 n\pi$ for integer $n\gg 1$ and $\tau \gg 0$, the atoms will reach population consensus as $t$ approaches $l\tau$ ($l=1,2,\cdots$) if $\gamma_{1R} = \gamma_{1L} = (N-1)\gamma_{jR} =(N-1)\gamma_{jL}$ with $j = 2,3,\cdots,N$.
\end{thm}

\newtheorem*{Proof}{Proof}
\begin{Proof}
When $\omega_a \tau = 2n\pi$, the imaginary part of Eq.~(\ref{con:OneExciationVector}) equals zero. Moreover, as $\gamma_{1R} = \gamma_{1L} = (N-1)\gamma_{jR} =(N-1)\gamma_{jL} \triangleq (N-1)g$ with $j = 2,3,\cdots,N$, we have $A_0 = -B_0$ in Eq.~(\ref{con:OneExciationVector}). 

In the following, we divide the evolution time of the atoms into separated time slices as $[0,\tau), [\tau,2\tau),\cdots, [(l-1)\tau, l \tau),\cdots$.
and we denote $x(t)$ in the $l$th slice as $x^{(l)}(t)$.

In the first time slice  $0 \leq t < \tau$, Eq.~(\ref{con:OneExciationVector}) evolves according to $\dot{x}(t) = A_0 x(t)$ under the initial condition $x(0) = [1,0,\cdots,0]^T$. Solving it we get amplitude vector $x^{(1)}(t)$ of the form
\begin{subequations} \label{con:c1cjtSolution}
\begin{numcases}{}
c_1^{(1)}(t) =\frac{1}{N} + \frac{N-1}{N}e^{-N(N-1)g^2 t}\triangleq \frac{1}{N} + (N-1)\epsilon_1(t) ,   \label{c1t0tau}\\
c_j^{(1)}(t) =-\frac{1}{N} +\frac{1}{N} e^{-N(N-1)g^2t} \triangleq -\frac{1}{N} + \epsilon_1(t), j=2,\cdots,N, \label{cjt0tau}
\end{numcases}
\end{subequations}
where $\epsilon_1(t) = \frac{1}{N}e^{-N(N-1)g^2t}  \leq \frac{1}{N}$ and  $\lim_{t\rightarrow \infty} \epsilon_1(t) = 0$. Because $n\gg 1$ and $\tau \gg 0$, we have $\lim_{t\rightarrow \tau} \epsilon_1(t) \approx 0$.

In the second time slice $\tau \leq t < 2\tau$,  from Eq.~(\ref{con:OneExciationVector})  we have
\begin{small}
\begin{equation} \label{con:Cjs0Tau}
\begin{aligned}
\dot{x}^{(2)}(t) &=- B_0 x^{(2)}(t) + B_0 \begin{bmatrix}
\frac{1}{N} + \frac{N-1}{N}e^{-N(N-1)g^2  (t -\tau)}\\
 -\frac{1}{N} +\frac{1}{N} e^{-N(N-1)g^2 (t -\tau)} \\
 \vdots \\
 -\frac{1}{N} +\frac{1}{N} e^{-N(N-1)g^2 (t -\tau)}
  \end{bmatrix}\\
 &= -B_0\left ( x^{(2)}(t) - \begin{bmatrix}
\frac{1}{N} \\
 -\frac{1}{N}\\
 \vdots \\
 -\frac{1}{N}
  \end{bmatrix}\right ) +\epsilon_1(t-\tau)B_0 \begin{bmatrix}
N-1 \\
 1\\
 \vdots \\
 1
  \end{bmatrix} \\
&=  -B_0 x^{(2)}(t)  +N\epsilon_1(t-\tau)B_0 \left [ 1, 0,\cdots,0\right ]^T,
\end{aligned}
\end{equation}
\end{small}%
where we have used the fact that $B_0 \left [ 1, -1,\cdots,-1\right ]^T = 0.$ Denote $\alpha = N(N-1)g^2$  to simplify notation in the following discussions. Let $u = t-\tau \in [0,\tau)$. Then from Eq. \eqref{con:Cjs0Tau} we have $\dot{x}^{(2)}(u+\tau) = -B_0 x^{(2)}(u+\tau)  +N\epsilon_1(u)B_0 \left [ 1, 0,\cdots,0\right ]^T$. Denote $\mathbf{X}_2(s)$ as the Laplace transform of $x^{(2)}(u)$ when $u \geq 0$. We have 
\begin{equation} \label{con:ChangeXLaplace}
\begin{aligned}
s\mathbf{X}_2(s) -e^{-s\tau} x^{(1)}(\tau) = -B_0 \mathbf{X}_2(s)  + \frac{e^{-s\tau}}{s+\alpha}B_0 \left [ 1, 0,\cdots,0\right ]^T ,
\end{aligned}
\end{equation}
which can be re-written as
\begin{equation} \label{con:ChangeXLaplace2}
\begin{aligned}
\mathbf{X}_2(s)  = \left (sI+B_0 \right)^{-1}e^{-s\tau}x^{(1)}(\tau)  + \left (sI+B_0 \right)^{-1}\frac{e^{-s\tau}}{s+\alpha}B_0 \left [ 1, 0,\cdots,0\right ]^T. 
\end{aligned}
\end{equation}
Thus, in the time slice $\tau \leq t < 2\tau$,
\begin{equation} \label{con:ChangeXLaplaceTime}
\begin{aligned}
x^{(2)}(t)  = x^{(1)}(t) + h^{(1)}(t-\tau),
\end{aligned}
\end{equation}
where  $h^{(1)}(t)$ is  the  inverse Laplace transform of $\frac{\left (sI+B_0 \right)^{-1}}{s+\alpha}B_0 [1,0,\cdots,0]^T$. Because $\alpha >0$, the second component of RHS of Eq.~(\ref{con:ChangeXLaplaceTime}) approaches zero when $t\rightarrow 2\tau$, similar to the case in Eq.~(\ref{con:c1cjtSolution}) when $t\rightarrow \tau$.

In the third time slice $2\tau \leq t < 3\tau$, we have 
\begin{equation} \label{con:X3t}
\begin{aligned}
\dot{x}^{(3)}(t) &= -B_0 x^{(3)}(t) + B_0 x^{(2)}(t-\tau)\\
& = B_0 \left [ -x^{(3)}(t) + x^{(1)}(t-\tau) + h^{(1)}(t-2\tau) \right ].
\end{aligned}
\end{equation}
Similarly, denote $v = t-2\tau \in [0, \tau)$, then we can get $\dot{x}^{(3)}(v+2\tau) = -B_0 x^{(3)}(v+2\tau) + B_0 x^{(1)}(v + \tau) + B_0h^{(1)}(v)$. Denote $\mathbf{X}_3(s)$ as the Laplace transform of $x^{(3)}(v)$ when $v \geq 0$, we have 
\begin{equation} \label{con:ChangeXLaplace2tau3tau}
\begin{aligned}
&~~~~s\mathbf{X}_3(s) -e^{-2s\tau}x^{(2)}(2\tau)\\ 
&= -B_0 \mathbf{X}_3(s) + B_0 \mathbf{X}_1(s)e^{-s\tau} + B_0 \frac{\left (sI-A_0 \right)^{-1}e^{-2s\tau}}{s+\alpha}B_0 [1,0,\cdots,0]^T,
\end{aligned}
\end{equation}
which can be rewritten as
\begin{footnotesize}
\begin{equation} \label{con:ChangeXLaplace2tau3tau2}
\begin{aligned}
&~~~~\mathbf{X}_3(s)  =  \left (sI-A_0 \right)^{-1}e^{-2s\tau}x^{(2)}(2\tau)  + \left (sI-A_0 \right)^{-1}B_0 \mathbf{X}_1(s)e^{-s\tau}\\
&~~~~+\left (sI-A_0 \right)^{-1}B_0 \frac{\left (sI-A_0 \right)^{-1}e^{-2s\tau}}{s+\alpha}B_0 [1,0,\cdots,0]^T\\
&=\left (sI-A_0 \right)^{-1}e^{-2s\tau}x^{(1)}(2\tau) + \left (sI-A_0 \right)^{-1}e^{-2s\tau}h^{(1)}(\tau)+ \left (sI-A_0 \right)^{-1}B_0 \mathbf{X}_1(s)e^{-s\tau}\\
&~~~~+\left (sI-A_0 \right)^{-1}B_0 \frac{\left (sI-A_0 \right)^{-1}e^{-2s\tau}}{s+\alpha}B_0 [1,0,\cdots,0]^T.
\end{aligned}
\end{equation}
\end{footnotesize}%
Similar to Eq.~(\ref{con:ChangeXLaplaceTime}), $\lim_{t\rightarrow 3\tau} x^{(3)}(t) \approx x^{(2)}(2\tau)$. The above process can be similarly generalized to  cases $t\in [(l-1)\tau, l \tau)$ for $l=4,5,\cdots$.
Thus $x^{(l)}(t)$ approaches to $\mathcal{L}^{-1}\left (sI-A_0 \right)^{-1} x(0)$ when $t \rightarrow l\tau$.
\end{Proof}

\begin{remark}
In \textbf{Theorem~\ref{consensus}}, $\tau\gg 0$ means that in each time slice $[(l-1)\tau, l\tau)$, the quantum state amplitudes approach   constant values as $t\to l\tau$, as analyzed in the proof. Then the quantum consensus can be realized when the condition in \textbf{Theorem~\ref{consensus}} is satisfied. 
\end{remark}

\subsubsection{An example of four atoms}
\begin{figure}[h]
\centerline{\includegraphics[width=1\columnwidth]{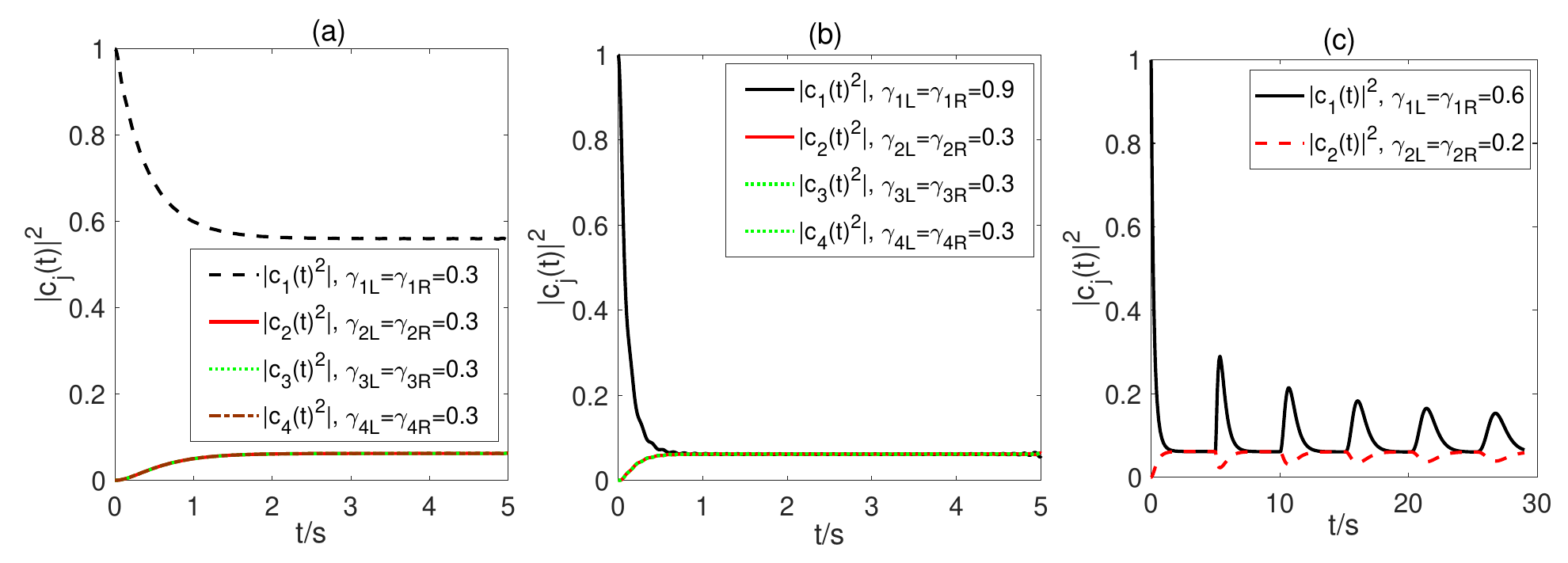}}
\caption{Evolution of four atoms coupled with a semi-infinite waveguide.}
	\label{fig:FouratomMultiPeriod}
\end{figure}

In this subsection, we use an example to demonstrate  \textbf{Theorem~\ref{consensus}}.   Consider four atoms that are located at $z_1 = 40\pi/\omega_a$ and nonchirally coupled to the waveguide. Suppose that initially only the first atom is excited, all the other three atoms are at the ground state, and the waveguide is empty.  Set $\gamma_{jR} = \gamma_{jL} = 0.3$ for $j=2,3,4$ and resonant frequency $\omega_a = 50$. We look at three scenarios. In scenario 1, $\gamma_{1R} = \gamma_{1L} = 0.3$. The first atom decays and the emitted field in the waveguide can be absorbed by the other three atoms with the same rate. Thus the amplitudes of the excited state of the second, third or fourth atom reach consensus, as shown in Fig.~\ref{fig:FouratomMultiPeriod}(a). In scenario 2,  $\gamma_{1R} = \gamma_{1L} = 0.9$. This is the setting in \textbf{Theorem~\ref{consensus}}. The populations of the excited states of the four individual atoms finally reach consensus, as shown  in Fig.~\ref{fig:FouratomMultiPeriod}(b).  In scenario 3,  $\gamma_{1R} = \gamma_{1L} = 3\gamma_{jR} = 3\gamma_{jL} = 0.6$ with $j=2,3,4$. This is also the setting in \textbf{Theorem~\ref{consensus}}.  The long time evolution is shown in Fig.~\ref{fig:FouratomMultiPeriod}(c). It can be seen that the long delay $\tau = 2z_1/c \approx 5.0265s$ can induce a perturbation on the evolution  of $|c_j(t)|^2$ when $t= (l-1)\tau$ with $l=2,3,\cdots$ and finally the atom network reach population consensus when $t\rightarrow l\tau$ as stated in \textbf{Theorem~\ref{consensus}}. The populations curves of $|c_3(t)|^2$ and $|c_4(t)|^2$ overlap with $|c_2(t)|^2$, thus are omitted.

\subsection{Coherent feedback control with two excitations}

When initially two atoms are excited, while   the waveguide is empty, there are  overall two excitations in the whole quantum system.
The quantum state in Fig.~\ref{fig:NatomWaveguide} (a) is 
\begin{equation} \label{con:StateTwoExcitation}
\begin{aligned}
|\Psi(t)\rangle =&\sum_{1\leq j < l\leq N}c_{jl}(t)|g_1,\cdots,e_j,\cdots,e_{l},\cdots,g_{N},\{0\}\rangle \\
&+ \sum_{j=1}^N \int c_j^{\bullet}(t,k)|g_1,\cdots,g_{j-1},e_{j},g_{j+1},\cdots, g_{N},\{k\}\rangle \\
&+ \int\int c_{*}(t,k_1,k_2) |g_1,\cdots, g_{N},\{k_1\}\{k_2\}\rangle \mathrm{d}k_1\mathrm{d}k_2, \\
\end{aligned}
\end{equation}
where the first component on the RHS represents that two atoms labeled $j$ and $l$ are excited and there are no photons in the waveguide, the second component means that the $j$th atom is excited and there is one photon of the continuous mode $k$ in the waveguide, and the last term indicates that all the atoms are in their ground states and there are two photons of the modes $k_1$ and $k_2$ in the waveguide, respectively. The quantum state representation can be mapped to a random graph as given in \textbf{Definition~\ref{GraphB}} below.

\bigskip

\begin{myDef} \label{GraphB}
The quantum coherent feedback system  in Fig.~\ref{fig:NatomWaveguide} (a) with two excitations can be mapped to the random graph in Fig.~\ref{fig:NatomWaveguide} (b). Specifically, the $j$th vertex represents that the $j$th two-level atom is excited and there is one photon in the semi-infinite waveguide; the edge between the $j$th and $l$th vertices means that the $j$th and $l$th two-level atoms are both excited and there are no photons in the waveguide. The probabilities of the $j$th vertex and the edge between the $j$th and $l$th vertices are
\begin{subequations} \label{con:GraphBRelation}
\begin{numcases}{}
p_j(t) = \int |c_j^{\bullet}(t,k)|^2\mathrm{d}k, \\
p_{jl}(t) = |c_{jl}(t)|^2.
\end{numcases}
\end{subequations}
\end{myDef}

Substituting the state representation in Eq.~(\ref{con:StateTwoExcitation}) into the Schr\"{o}dinger equation,  we can derive the evolution of the amplitudes as
\begin{small}
\begin{subequations} \label{con:Popuquation}
\begin{numcases}{}
\dot{c}_{jl}(t) = - i\int c_{j}^{\bullet}(t,k)g_{klt}^*(k,t,z_l)\mathrm{d}k -i \int c_{l}^{\bullet}(t,k) g_{kjt}^*(k,t,z_j)\mathrm{d}k,\, l\neq j,\label{NAmodel1}\\
\dot{c}_{j}^{\bullet}(t,k) = -i \sum_l c_{jl}(t)  g_{klt}(k,t,z_l)  - i  \int c_{*}(k,k_2,t)g_{kjt}^*(k_2,t,z_j) \mathrm{d}k_2,\label{NAmodel2}\\
\dot{c}_{*}(k_1,k_2,t) = -i \sum_j c_{j}^{\bullet}(t,k_1)  g_{kjt}(k_2,t,z_j)  -i \sum_j c_{j}^{\bullet}(t,k_2)  g_{kjt}(k_1,t,z_j). \label{NAmodel3}
\end{numcases}
\end{subequations}
\end{small}%
 Eq.~(\ref{NAmodel1}) means that when  one of the $N$ atoms is excited and there is one photon in the waveguide, then another atom can also be excited after absorbing the photon from the waveguide. The first component of the RHS of Eq.~(\ref{NAmodel2}) means that when initially the two atoms are both excited and there are no photons in the waveguide, then any of these two atoms can emit one photon into the waveguide; the second component represents that an atom can also absorb one photon from the waveguide if there are two photons in the waveguide. Eq.~(\ref{NAmodel3}) means that when there is one photon in the waveguide and one of the atoms is excited, then the excited atom can emit one photon into the waveguide to generate a two-photon state.

According to the calculations in Appendix~\ref{Sec:ApendixDelay}, Eqs.~(\ref{NAmodel1},\ref{NAmodel2}) can be re-written in the delay dependent format as
\begin{scriptsize}
\begin{subequations}\label{con:PopuquationDelay}
\begin{align}
&\dot{c}_{jl}(t) = - \frac{\gamma_{jR}^2 + \gamma_{jL}^2+\gamma_{lR}^2 + \gamma_{lL}^2}{2} c_{jl}(t) + \gamma_{lL} \gamma_{lR} c_{jl}\left(t-\frac{2z_l}{c}\right)e^{i\omega_a\frac{2z_l}{c}}+ \gamma_{jL} \gamma_{jR} c_{jl}\left(t-\frac{2z_j}{c}\right)e^{i\omega_a\frac{2z_j}{c}} \notag\\
&-\sum_{l'=1,l'> j}^{l-1}\gamma_{l'R}\gamma_{lR}c_{jl'}\left(t-\frac{z_l-z_{l'}}{c}\right) e^{i\omega_a \frac{z_l-z_{l'}}{c}}-\sum_{l'=l+1,l'> j}^{N}\gamma_{l'L}\gamma_{lL}c_{jl'}\left (t-\frac{z_{l'}-z_l}{c}\right )e^{i\omega_a \frac{z_{l'}-z_l}{c}} \notag\\
&-\sum_{j'=1,j'< l}^{j-1}\gamma_{j'R}\gamma_{jR}c_{j'l}\left(t-\frac{z_j-z_{j'}}{c}\right) e^{i\omega_a\frac{z_j-z_{j'}}{c}} -\sum_{j'=j+1,j'< l}^{N}\gamma_{j'L}\gamma_{jL}c_{j'l}\left(t-\frac{z_{j'}-z_j}{c}\right)e^{i\omega_a \frac{z_{j'}-z_j}{c}} \notag\\
&+ \sum_{l'\neq l} \gamma_{l'L}\gamma_{lR}c_{jl}\left(t-\frac{z_l+z_{l'}}{c}\right)e^{i\omega_a \frac{z_l+z_{l'}}{c}} + \sum_{j'\neq j} \gamma_{j'L}\gamma_{jR}c_{jl}\left(t-\frac{z_j+z_{j'}}{c}\right) e^{i\omega_a \frac{z_j+z_{j'}}{c}},\label{NAmodel1Delay}\\
&\dot{c}_{j}^{\bullet}(t,k) =  - \frac{\gamma_{jR}^2 + \gamma_{jl}^2}{2}c_{j}^{\bullet}(t,k) -i \sum_l c_{jl}(t)  g_{klt}(k,t,z_l)  + \sum_{j'}\gamma_{j'R} \gamma_{jL}c_{j}^{\bullet}\left(t-\frac{z_j+z_{j'}}{c},k\right ) e^{i\omega_a\frac{z_j+z_{j'}}{c}} \notag\\
&-\sum_{j'<j}\gamma_{j'R} \gamma_{jR} c_j^{\bullet}\left(t-\frac{z_j-z_{j'}}{c}\right) e^{i\omega_a\frac{z_j-z_{j'}}{c}}-\sum_{j'>j}\gamma_{j'L}\gamma_{jL} c_{j}^{\bullet}\left(t-\frac{z_{j'}-z_j}{c},k\right)e^{i\omega_a\frac{z_{j'}-z_j}{c}}.\label{NAmodel2Delay}
\end{align}
\end{subequations}
\end{scriptsize}%
The first component of RHS of Eq.~(\ref{NAmodel1Delay}) shows the spontaneous emission process of the two excited atoms into the waveguide, and the next two components represent that one atom can emit one photon into the waveguide and the photon can be further reflected by the terminal mirror to induce the round-trip transmission delay. The terms on the RHS of the second and third lines of Eq.~(\ref{NAmodel1Delay}) represent the exchange of excited states between an excited atom with one of the other atoms at the ground state and this process is not affected by the mirror reflection.  The  components on the RHS of the fourth line of Eq.~(\ref{NAmodel1Delay})  indicate the excitation is transferred from one atom to another atom after the reflection by the mirror. Compared with Eq.~(\ref{NAmodel2}), the first component on the RHS of Eq.~(\ref{NAmodel2Delay}) gives the spontaneous emission of the excited atom, the second item represents that one of the excited atoms can emit one photon into the waveguide, and the following three components with delays represent the transmission of the excited state from one atom to another atom at the ground state.

When there are two atoms (namely $N=2$) coupled with the semi-infinite waveguide, the atoms can be continuously excited, as introduced in~\cite{DingZhangAutomatica} with the following theorem.

\begin{thm}
[\cite{DingZhangAutomatica}] When $N=2$, $\gamma_{1R} = \gamma_{1L}$, $\gamma_{2R} = \gamma_{2L}$, $z_1 = z_2 = n\pi/\omega_a$, where $n=1,2,\cdots$, and $c_{12}(0) = 1$, then the two atoms can be continuously excited.
\end{thm}

For the general $N>2$ case, we consider the special scenario that $z_1 = z_2 = \cdots = z_N$ with $n\gg 1$, similar to the settings in \textbf{Theorem~\ref{consensus}}. In this case, when $t<\tau$, Eq.~(\ref{con:PopuquationDelay}) reduces to
\begin{scriptsize}
\begin{subequations}\label{con:TwoexcitationLargetau}
\begin{align}
\dot{c}_{jl}(t) = & - \frac{\gamma_{jR}^2 + \gamma_{jL}^2+\gamma_{lR}^2 + \gamma_{lL}^2}{2} c_{jl}(t) 
-\sum_{l'=1,l'> j}^{l-1}\gamma_{l'R}\gamma_{lR}c_{jl'}\left(t\right) -\sum_{l'=l+1,l'> j}^{N}\gamma_{l'L}\gamma_{lL}c_{jl'}\left (t \right ) \notag\\
&-\sum_{j'=1,j'< l}^{j-1}\gamma_{j'R}\gamma_{jR}c_{j'l}\left(t\right)  -\sum_{j'=j+1,j'< l}^{N}\gamma_{j'L}\gamma_{jL}c_{j'l}\left(t\right),\label{TwoExcLongdelay1}\\
\dot{c}_{j}^{\bullet}(t,k) =&- \frac{\gamma_{jR}^2 + \gamma_{jl}^2}{2}c_{j}^{\bullet}(t,k) -i \sum_l c_{jl}(t)  g_{klt}(k,t,z_l)   -\sum_{j'<j}\gamma_{j'R} \gamma_{jR} c_j^{\bullet}\left(t,k\right) \notag\\
&-\sum_{j'>j}\gamma_{j'L}\gamma_{jL} c_{j}^{\bullet}\left(t,k\right).\label{TwoExcLongdelay2}
\end{align}
\end{subequations}
\end{scriptsize}

\begin{figure}[h]
\centerline{\includegraphics[width=1\columnwidth]{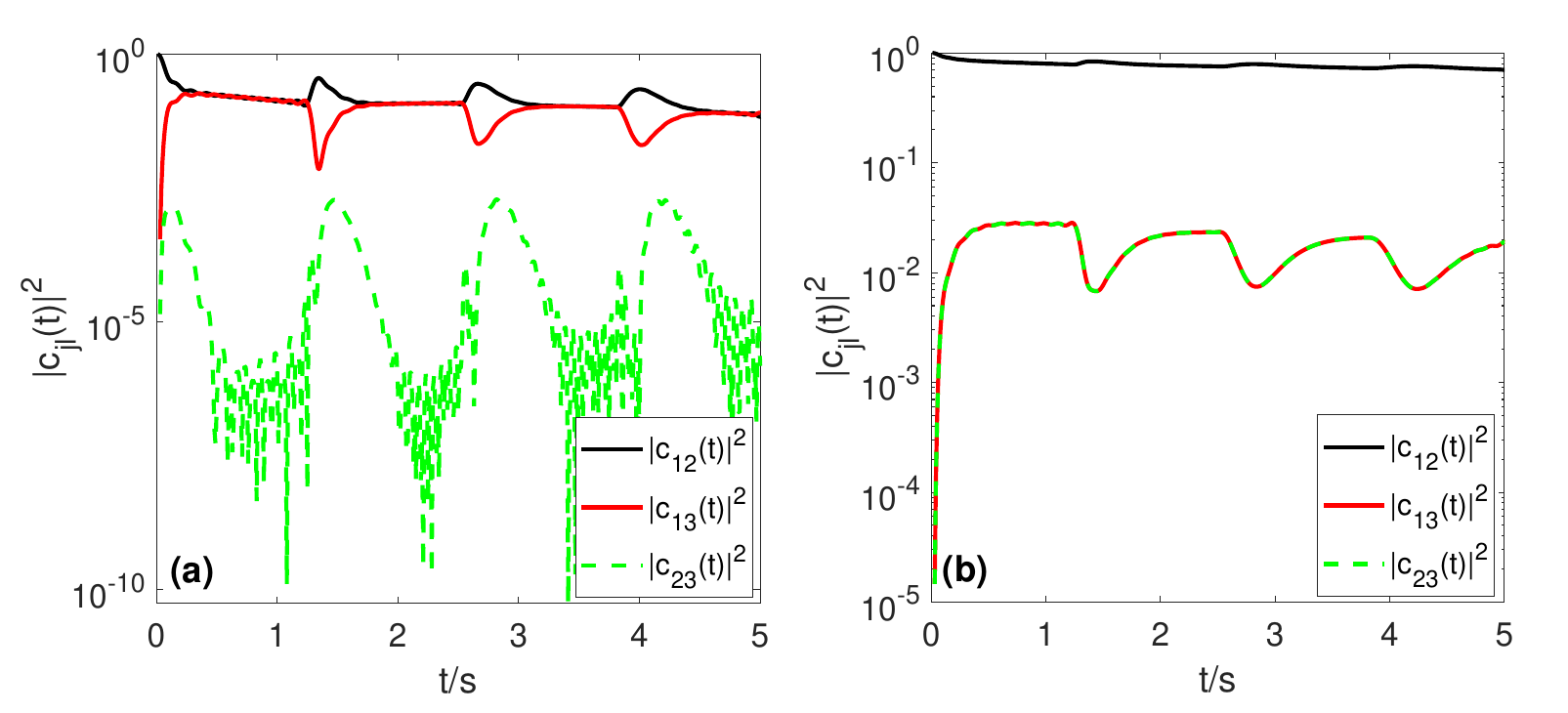}}
\caption{Consensus evolution of three atoms with two excitations coupled with a semi-infinite waveguide.}
	\label{fig:FouratomTwoexcitation}
\end{figure}

Eq.~(\ref{TwoExcLongdelay1}) indicates that the amplitude representing two excited atoms oscillates according to the atom-waveguide couplings. This can be interpreted in terms of the probability of the edges of a random graph. As illustrated in Fig.~\ref{fig:FouratomTwoexcitation} with $N=3$, there are three two-level atoms at $z_j = 40\pi/\omega_a$ with $\omega_a = 50$ and $\tau \approx 5.0265s$.  Initially only the first two atoms are excited, i.e., $c_{12}(0) =1$. In Fig.~\ref{fig:FouratomTwoexcitation} (a), $\gamma_{1R} = \gamma_{1L} = 0.2$, $\gamma_{2R} = \gamma_{2L} =\gamma_{3R} = \gamma_{3L} = 1$. When $t<\tau$, both the first and second atoms can emit photons into the waveguide, which can be absorbed by the third atom. Thus  the first and the third atoms, or the second and the third atoms can be simultaneously excited.  In Fig.~\ref{fig:FouratomTwoexcitation} (b), $\gamma_{1R} = \gamma_{1L} = \gamma_{2R} = \gamma_{2L} = 0.2$, $\gamma_{3R} = \gamma_{3L} = 1$, $c_{13}$ and $c_{23}$ evolve with the same parameter settings according to Eq.~(\ref{TwoExcLongdelay1}) and can reach population consensus. 

In summary, in this section we have investigated the coherent feedback dynamics of a quantum network where multiple two-level atoms are coupled with a semi-infinite waveguide, with either one or two atoms initially being in the excited state. 

\subsection{The coherent feedback dynamics under random couplings among atoms}
In this section, we study how the the random couplings among atoms can affect the coherent feedback dynamics when $z_j = z_l$ for arbitrary $j$ and $l$, i.e., all atoms are collocated. Based on Eq.~(\ref{con:Hintj2nonchiral}), the Hamiltonian in the interaction picture with random couplings reads~\cite{ruostekoski2017arrays,whitlock2017simulating}
\begin{small}
\begin{equation} \label{con:HintjRandomCoupling}
\begin{aligned}
H_R &= \sum_j H_{I}^{(j)} + \sum_{j\neq l} \delta_{jl} \left ( \sigma_j^- \sigma_l^+ + \sigma_l^-\sigma_j^+\right ),
\end{aligned}
\end{equation}
\end{small}%
where $H_{I}^{(j)}$ is given in Eq.~(\ref{con:Hintj2nonchiral}), and $\left|\delta_{jl}\right| \leq \Delta$ is the random coupling between the $j$th and $l$th atoms. Based on this, we study the  dynamics of the quantum coherent feedback network in the presence of one or  two excitations.

\subsubsection{One excitation with random couplings}
\begin{figure}[h]
\centerline{\includegraphics[width=1.1\columnwidth]{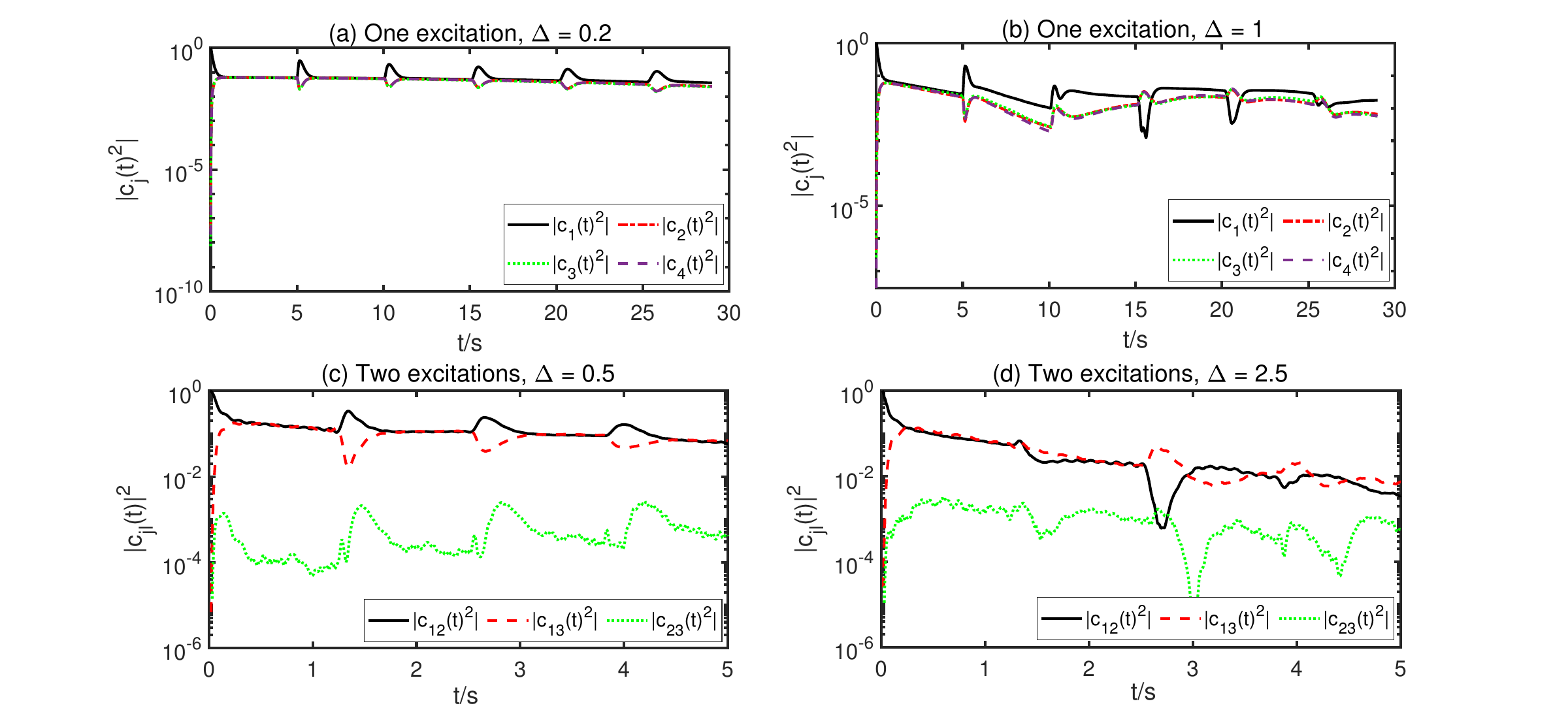}}
\caption{The consensus realization influenced by random couplings.}
	\label{fig:RandomCoupling}
\end{figure}

When initially only the first atom is excited and there are no photons in the waveguide, the quantum state can still be formally represented as in Eq.~(\ref{con:State}).  However, to account for the effect of the random couplings, we have to replace the amplitudes $c_j(t)$ and $\tilde{c}(t,k)$ with $\hat{c}_j(t)$ and $\hat{\tilde{c}}(t,k)$, respectively. Generalized from Eq.~(\ref{con:WavNatom}), the evolution of quantum state amplitudes can be written as
\begin{footnotesize}
\begin{subequations} \label{con:WavNatomRandomCoupling}
\begin{numcases}{}
\dot{\hat{c}}_j(t) =- i \sum_{l \neq j} \delta_{jl} \hat{c}_l(t) \notag\\
~~~~~~~~~~+ \int \left \{ -\gamma_{jR} e^{-i[(\omega_k-\omega_a)t - \omega_k z_j/c]} +\gamma_{jL}e^{-i[(\omega_k-\omega_a)t + \omega_k z_j/c]}\right\}\hat{\tilde{c}}(t,k) \mathrm{d}k,   \label{cjtRandCou}\\
\dot{\tilde{c}}(t,k) 
= \sum_{j=1}^N  \hat{c}_j(t) \left \{ \gamma_{jR} e^{i[(\omega_k-\omega_a)t - \omega_k z_j/c]} - \gamma_{jL}e^{i[(\omega_k-\omega_a)t + \omega_k z_j/c]}\right\}, \label{cktRandCou}
\end{numcases}
\end{subequations}
\end{footnotesize}%
where the first component on the RHS of Eq.~(\ref{cjtRandCou}) is induced by the random couplings between arbitrary two atoms. The influence of random couplings on the consensus realization is demonstrated in Fig.~\ref{fig:RandomCoupling} with $\Delta = 0.2$ in (a), $\Delta = 1$ in (b), and the other parameter settings are identical to those in Fig.~\ref{fig:FouratomMultiPeriod}(b).
The comparisons show that the consensus in the atom network still  exists as long as the random coupling strengths among the atoms are small.

\subsubsection{Two excitations with random couplings}
For the case that there are two excitations in the quantum network, we assume initially the first two atoms are excited and there are no photons in the waveguide. The Hamiltonian of the atom network with random couplings are identical to that in Eq.~(\ref{con:HintjRandomCoupling}). However, as there are two excitations, we need to replace the amplitudes $c_{jl}(t)$, $c_j^{\bullet}(t,k)$ and $c_{*}(t,k_1,k_2)$ in Eq.~(\ref{con:TwoexcitationLargetau}) with $\hat{c}_{jl}(t)$, $\hat{c}_j^{\bullet}(t,k)$ and $ \hat{c}_{*}(t,k_1,k_2)$, respectively. Then the evolutions of the quantum state amplitudes read
\begin{scriptsize}
\begin{subequations}\label{con:TwoexcitationLargetauRandomCoupling}
\begin{align}
\dot{\hat{c}}_{jl}(t) = & - \frac{\gamma_{jR}^2 + \gamma_{jL}^2+\gamma_{lR}^2 + \gamma_{lL}^2}{2} \hat{c}_{jl}(t) 
-\sum_{l'=1,l'> j}^{l-1}\gamma_{l'R}\gamma_{lR}\hat{c}_{jl'}\left(t\right) -\sum_{l'=l+1,l'> j}^{N}\gamma_{l'L}\gamma_{lL}\hat{c}_{jl'}\left (t \right ) \notag\\
&-\sum_{j'=1,j'< l}^{j-1}\gamma_{j'R}\gamma_{jR}\hat{c}_{j'l}\left(t\right)  -\sum_{j'=j+1,j'< l}^{N}\gamma_{j'L}\gamma_{jL}\hat{c}_{j'l}\left(t\right)\notag\\
&-i\sum_{j' \neq j, j'<l}\delta_{jj'}  \hat{c}_{j'l}(t) - i\sum_{l' \neq l, l'>j}\delta_{ll'}\hat{c}_{jl'}(t), \label{TwoExcLongdelay1RM}\\
\dot{\hat{c}}_{j}^{\bullet}(t,k) =&- \frac{\gamma_{jR}^2 + \gamma_{jl}^2}{2}\hat{c}_{j}^{\bullet}(t,k) -i \sum_l \hat{c}_{jl}(t)  g_{klt}(k,t,z_l)   -\sum_{j'<j}\gamma_{j'R} \gamma_{jR} \hat{c}_j^{\bullet}\left(t,k\right) \notag\\
&-\sum_{j'>j}\gamma_{j'L}\gamma_{jL} \hat{c}_{j}^{\bullet}\left(t,k\right) 
 - i \sum_{l \neq j} \delta_{jl} \hat{c}_{l}^{\bullet}(t,k),\label{TwoExcLongdelay2RM}\\
 \dot{\hat{c}}_{*}&(k_1,k_2,t) = -i \sum_j \hat{c}_{j}^{\bullet}(t,k_1)  g_{kjt}(k_2,t,z_j)  -i \sum_j \hat{c}_{j}^{\bullet}(t,k_2)  g_{kjt}(k_1,t,z_j),\label{TwoExcLongdelay3RMHat}
\end{align}
\end{subequations}
\end{scriptsize}%
where the last two components of Eq.~(\ref{TwoExcLongdelay1RM}) and the last component of Eq.~(\ref{TwoExcLongdelay2RM}) represent how the quantum state amplitudes are influenced by the random couplings between arbitrary two atoms, and Eq.~(\ref{TwoExcLongdelay3RMHat}) has the same meaning as Eq.~(\ref{NAmodel3}). The influence by random couplings when there are two excitations are compared in Fig.~\ref{fig:RandomCoupling} with $\Delta = 0.5$ in (c), $\Delta = 2.5$ in (d), and the other parameters settings are the same as those in Fig.~\ref{fig:FouratomTwoexcitation}(a). Again, the atom network consensus can be maintained when the random coupling strength is small.

Apart from the random couplings among atoms, the quantum coherent feedback dynamics can also be influenced by the environment, which will be analyzed in the following section by combining with the quantum measurement feedback.

\section{Hybrid coherent and measurement feedback control} \label{sec:hybridopen}
In practice, the quantum system in Fig.~\ref{fig:NatomWaveguide} can be affected by the environment. For example, an initially excited atom  cannot be perfectly coupled with the waveguide, and thus  spontaneously decays to the (unmodelled) environment. As a result, the conclusions in Section~\ref{sec:coherentgraph} cannot perfectly hold. To compensate the influence of the environment, external drives must be applied to the atoms to generate desired quantum states. Take the case that there is only one two-level atom which is located at $z_1$ and has the resonant frequency $\omega_{a1}$. The Hamiltonian of the system with a coherent drive on the atom reads~\cite{pichler2015quantum,yan2019utilizing}
\begin{equation} \label{con:HtotalOneatom}
\begin{aligned}
H_t = & \left ( \omega_{a1} -i\gamma_0\right ) \sigma_1^+ \sigma_1^- + \frac{1}{2}\left (\Omega_1 \sigma_1^- + \Omega_1^* \sigma_1^+ \right ) + \int_0^{\infty}  \omega_k d^{\dag}_kd_k \mathrm{d}k \\
&+  \int  \left[g_{k1t}(k,t,z_1) d^{\dag}_k\sigma_1^-  + g_{k1t}^*(k,t,z_1)d_k\sigma_1^+ \right] \mathrm{d}k,\\
\end{aligned}
\end{equation}
where $\gamma_0$ is the atom's spontaneous decaying rate to the unmodelled environment, $\Omega_1 = \bar{\Omega}_1 e^{i\bar{\omega}_1t}$ represents a coherent drive with Rabi frequency $\bar{\Omega}_1$ at  frequency $\bar{\omega}_1$.  We denote the detuning between the coherent drive and the atom as $\delta_1 =\bar{\omega}_1 -  \omega_{a1}$.  Then  the atomic Hamiltonian in a rotating frame is~\cite{pichler2015quantum,Zoller2016PRL}
\begin{equation} \label{con:SystemHam}
\begin{aligned}
H_s = -\left (\delta_1 + i\gamma_0\right ) \sigma_1^+ \sigma_1^- + \frac{1}{2}\left ( \bar{\Omega}_1 \sigma_1^- + \rm{H.c.}\right).
\end{aligned}
\end{equation}

In the following discussion, we first take $\gamma_0 = 0$ to derive the quantum control dynamics, after that we analyze how a non-zero $\gamma_0$ influences the performance. Tracing out the waveguide yields the Lindblad master equation for the atom ~\cite{Anton2018PRL,ask2022non,valente2016non,fang2018non,bound1}
\begin{equation} \label{con:masterDissipation}
\begin{aligned}
\dot{\rho}(t) 
&= -i\left [H_{\mathrm{eff}},\rho(t) \right]  + \Gamma_{\mathrm{eff}} \left (\sigma_1^- \rho(t)\sigma_1^+ -\frac{1}{2} \rho(t)\sigma_1^+\sigma_1^- - \frac{1}{2}\sigma_1^+  \sigma_1^-\rho(t) \right )\\
&\triangleq  -i\left [H_{\mathrm{eff}},\rho(t) \right]  + \Gamma_{\mathrm{eff}} \mathcal{L}_{\sigma_1^-} \left [ \rho(t) \right ],
\end{aligned}
\end{equation}
where
\[
\mathcal{L}_{\sigma_1^-} \left [ \rho(t) \right ] = \sigma_1^- \rho(t)\sigma_1^+ -\frac{1}{2} \rho(t)\sigma_1^+\sigma_1^- - \frac{1}{2}\sigma_1^+  \sigma_1^-\rho(t) ,
\]
and the effective Hamiltonian is
\[
H_{\mathrm{eff}} = -\delta_1 \sigma_1^+ \sigma_1^- + \frac{1}{2}\left ( \bar{\Omega}_1 \sigma_1^- + \rm{H.c.}\right) - \gamma_{1L}\gamma_{1R}\sin\left(\omega_a \frac{2z_1}{c}\right)\sigma_1^+  \sigma_1^- , 
\]
and 
$\Gamma_{\mathrm{eff}} = y_{\gamma} + \Gamma$ with $y_{\gamma} = \left(\gamma_{1R}^2+\gamma_{1L}^2\right)/2  -\gamma_{1L}\gamma_{1R}\cos\left(2\omega_a z_1/c \right)$. 
Here $y_{\gamma}$ represents the dissipation rate from the atom to the waveguide, and $\Gamma\geq 0$ is the dissipation of the waveguide mode to the environment, which can be induced by the practical imperfect design such as that the atom is not at the central of the waveguide as in~\cite{WZH2022}. Clearly,  $\Gamma_{\mathrm{eff}} \geq 0$ where the equality holds only when $\Gamma = 0$, $\gamma_{1R} = \gamma_{1L},$ and  $\cos\left(2\omega_a z_1/c\right) = 1$.
\begin{remark}
When initially the atom is excited, $\delta_1  = \bar{\Omega}_1 =\Gamma = 0$, $\gamma_{1R} = \gamma_{1L}$ and $\omega_a z_1/c = n\pi$ with $n =1,2,\cdots$, then $\dot{\rho}(t) = 0$. In other words, the atom is continuously excited, as studied in Refs.~\cite{bradford2013spontaneous,DingZhangAutomatica,tufarelli2014non,barkemeyer2021strongly,bound1,dorner2002laser}.
\end{remark}

Denote $X =  \sigma_1^-+\sigma_1^+$, and $Y = \delta_1 + \gamma_{1L}\gamma_{1R}\sin\left(2\omega_a z_1/c\right)$.
Then from the Lindblad master equation \eqref{con:masterDissipation} we get the average values
\begin{small}
\begin{subequations} \label{con:OneatomMeanOperator}
\begin{numcases}{}
\frac{\mathrm{d}}{\mathrm{d}t} \langle X \rangle = i Y \left ( \langle \sigma_1^-\rangle -\langle \sigma_1^+\rangle \right )-\frac{\Gamma_{\mathrm{eff}}}{2}\left ( \langle \sigma_1^-\rangle +\langle \sigma_1^+\rangle \right ),   \label{OneAtomMean1}\\
\frac{\mathrm{d}}{\mathrm{d}t} \langle \sigma_1^+ \rangle = -i Y\langle \sigma_1^+\rangle -\frac{\Gamma_{\mathrm{eff}}}{2}\langle \sigma_1^+\rangle - \frac{i\bar{\Omega}}{2}\langle \sigma_1^z \rangle,   \label{OneAtomMean2}\\
\frac{\mathrm{d}}{\mathrm{d}t} \langle \sigma_1^- \rangle = i Y\langle \sigma_1^-\rangle -\frac{\Gamma_{\mathrm{eff}}}{2}\langle \sigma_1^-\rangle + \frac{i\bar{\Omega}}{2}\langle \sigma_1^z \rangle,   \label{OneAtomMean3}\\
\frac{\mathrm{d}}{\mathrm{d}t} \langle \sigma_1^z \rangle = -i\bar{\Omega} \left (\langle \sigma_1^+ \rangle -\langle \sigma_1^- \rangle \right) -\Gamma_{\mathrm{eff}} \left (\langle \sigma_1^z \rangle +1 \right),   \label{OneAtomMean4}
\end{numcases}
\end{subequations}
\end{small}%
where for arbitrary operator $\textbf{O}$, $\left \langle \textbf{O} \right \rangle  = \mathrm {Tr} \left[ \textbf{O} \rho\right]$. Let $\tilde{X} = \left [\langle \sigma_1^+ \rangle, \langle \sigma_1^- \rangle, \langle \sigma_1^z \rangle\right ]^\mathrm{T}$ with the superscript $\mathrm{T}$ denoting the  matrix transpose. Then from Eq. \eqref{con:OneatomMeanOperator} we have
\begin{equation} \label{con:meanOperatorEquation}
\begin{aligned}
\dot{\tilde{X}} &= \begin{bmatrix}
   -i Y-\frac{\Gamma_{\mathrm{eff}}}{2} & 0 &-\frac{i\bar{\Omega}}{2}\\
   0 &  iY - \frac{\Gamma_{\mathrm{eff}}}{2} &   \frac{i\bar{\Omega}}{2}\\
   -i\bar{\Omega} & i\bar{\Omega} & -\Gamma_{\mathrm{eff}}
  \end{bmatrix}  \tilde{X} -  \begin{bmatrix}
0\\ 0 \\ \Gamma_{\mathrm{eff}}
  \end{bmatrix}\\
  &\triangleq A\tilde{X} - B,
\end{aligned}
\end{equation}
and 
\begin{equation} \label{con:measurementOutput}
\begin{aligned}
\langle X \rangle &= \begin{bmatrix}
  1 & 1 &0
  \end{bmatrix}  \tilde{X} &\triangleq C\tilde{X}.
\end{aligned}
\end{equation}

Denote $\tilde{\textbf{X}}(s)$ as the Laplace transform of $\tilde{X}(t)$,  and $\textbf{X}(s)$ as that of $\langle X(t) \rangle$. Then from the above two equations we get
\begin{equation} \label{con:XLaplace}
\begin{aligned}
s\tilde{\textbf{X}}(s) - \tilde{X}(0) = A\tilde{\textbf{X}}(s) -\frac{1}{s}B,
\end{aligned}
\end{equation}
which is
\begin{equation} \label{con:XLaplace2}
\begin{aligned}
\tilde{\textbf{X}}(s)  = \left (sI-A \right)^{-1} \left ( \tilde{X}(0)-\frac{1}{s}B \right),
\end{aligned}
\end{equation}
and
\begin{equation} \label{con:XLaplace3}
\begin{aligned}
\textbf{X}(s)  = C\left (sI-A \right)^{-1} \left ( \tilde{X}(0)-\frac{1}{s}B \right).
\end{aligned}
\end{equation}

It can be readily verified that
\begin{equation} \label{con:SIminA}
\begin{aligned}
|sI-A| &=   \left [ \left ( s + \frac{\Gamma_{\mathrm{eff}}}{2}\right )^2 + Y^2 \right ] \left ( s + \Gamma_{\mathrm{eff}}\right ) + \bar{\Omega}^2 \left (s + \frac{\Gamma_{\mathrm{eff}}}{2} \right ).
\end{aligned}
\end{equation}

\begin{remark} \label{RemarkonStability}
When $\Gamma_{\mathrm{eff}} > 0$,   it is clear to see from Eq. \eqref{con:SIminA} that the two roots of the equation $|sI-A| = 0$ have negative real parts; thus  the linear control system in Eq.~(\ref{con:meanOperatorEquation}) is Hurwitz stable. This applies to all the following theorems (\textbf{Theorems \ref{Theorem3Finalstate}, \ref{Theorem5Y0}, \ref{Theorem6MF}, \ref{Cohestability}, \ref{transition}}), thus the steady states in all these theorems are asymptotically stable.
\end{remark}

Denote $Z(s)$ as the Laplace transform of $\langle \sigma_1^z(t) \rangle$. We analyze the atomic dynamics based on the following assumption. 

\bigskip

\begin{assumption} \label{AsumInitialOneatomDrive}
Assume initially $\tilde{X}(0) = \left [0, 0, z_0\right ]^\mathrm{T}$ with $-1 \leq z_0 \leq 1$.
\end{assumption}
Then by Eq. \eqref{con:XLaplace2}, we can get
\begin{equation} \label{con:ZsGeneral}
\begin{aligned}
Z(s)& = \frac{\left ( z_0 - \frac{\Gamma_{\mathrm{eff}}}{s}\right )\left [ \left (s+ \frac{\Gamma_{\mathrm{eff}}}{2}\right )^2+Y^2\right ] }{\left [ \left ( s + \frac{\Gamma_{\mathrm{eff}}}{2}\right )^2 + Y^2 \right ] \left ( s + \Gamma_{\mathrm{eff}}\right ) + \bar{\Omega}^2 \left (s + \frac{\Gamma_{\mathrm{eff}}}{2} \right )}.
\end{aligned}
\end{equation}

\begin{thm} \label{Theorem3Finalstate}
If $\Gamma > 0$, then the steady value of $\langle \sigma_1^z (\infty) \rangle <0$.
\end{thm}

\begin{Proof}
According to Eq.~(\ref{con:ZsGeneral}), the steady value of $\langle \sigma_1^z\rangle $ reads
\begin{equation} \label{con:Zsteadystate}
\begin{aligned}
\langle \sigma_1^z (\infty)\rangle & = \lim_{s\rightarrow 0} \frac{s\left ( z_0 - \frac{\Gamma_{\mathrm{eff}}}{s}\right )\left [ \left (s+ \frac{\Gamma_{\mathrm{eff}}}{2}\right )^2+Y^2\right ] }{\left [ \left ( s + \frac{\Gamma_{\mathrm{eff}}}{2}\right )^2 + Y^2 \right ] \left ( s + \Gamma_{\mathrm{eff}}\right ) + \bar{\Omega}^2 \left (s + \frac{\Gamma_{\mathrm{eff}}}{2} \right )}\\
& = -\frac{\Gamma_{\mathrm{eff}}^2 + 4Y^2}{\Gamma_{\mathrm{eff}}^2+ 4Y^2 + 2\bar{\Omega}^2} < 0.
\end{aligned}
\end{equation}
\end{Proof}

The conclusion above can also be interpreted in terms of the following theorem.

\begin{thm}[\cite{qi2012TAC}]\label{QiTAC}
 Consider a quantum control system with the  Hamiltonian $H_0$ and Lindblad coupling operator $L$. If $\left [ H_0,L\right ] \neq 0$, then there is at least one eigenstate of $H_0$ denoted as $\rho_d$, such that the trace distance $D\left ( \rho(t), \rho_d \right)$ between the quantum state $\rho(t)$ and $\rho_d$  satisfies that
\[
\limsup_{t\rightarrow\infty} D\left ( \rho(t), \rho_d \right) >0.
\]

\end{thm}
\begin{remark} \label{QiRemark}
\textbf{Theorem}~\ref{QiTAC} means that when the dissipation operator does not commute with the atomic Hamiltonian, there is at least one eigenstate which cannot be prepared with arbitrary high fidelity no matter how the external control is designed. In our case $L = \sigma_1^-$ does not commute with the atomic Hamiltonian $H_s$,  thus the atom cannot be continuously excited when the waveguide dissipates to the environment with the rate $\Gamma>0$ in Eq.~(\ref{con:masterDissipation}). This also agrees with the conclusion in Ref.~\cite{WZH2022} that $\Gamma$ can provide a new dissipation channel to prevent the atom's excitation. 
\end{remark}

\subsection{An example when $\gamma_0 \neq 0$ and $\Gamma = 0$}
When $\gamma_0 \neq 0$ and $\Gamma = 0$, denote $Y' = Y + i\gamma_0$ and $\Gamma_{\mathrm{eff}}' =  y_{\gamma}$.   Let the initial atomic state be that given  in \textbf{Assumption}~\ref{AsumInitialOneatomDrive}.  The steady state of the atom can be modified according to Eq.~(\ref{con:ZsGeneral}) as
\begin{equation} \label{con:ZsBoundDecay}
\begin{aligned}
Z'(s)& = \frac{\left ( z_0 - \frac{\Gamma_{\mathrm{eff}}'}{s}\right )\left [ \left (s+ \frac{\Gamma_{\mathrm{eff}}'}{2}\right )^2+Y'^2\right ] }{\left [ \left ( s + \frac{\Gamma_{\mathrm{eff}}'}{2}\right )^2 + Y'^2 \right ] \left ( s + \Gamma_{\mathrm{eff}}'\right ) + \bar{\Omega}^2 \left (s + \frac{\Gamma_{\mathrm{eff}}'}{2} \right )}.
\end{aligned}
\end{equation}

\begin{thm} \label{Theorem5Y0}
When $Y =  0$, the atom can be continuously excited and the population of the excited state is determined by the coherent drive and decaying to the environment. In particular, if $\bar{\Omega}^2 = 4\gamma_0^2 - \Gamma_{\mathrm{eff}}'^2$, then $\langle \sigma_1^z (\infty)\rangle = 1$, i.e., the coherent drive can 
compensate the atom's spontaneous decay and  thus the atom remains in the excited state.
\end{thm}
\begin{Proof}
The atomic population is given by $\langle \sigma_1^z \rangle$. When $Y=0$, $Y' = i\gamma_0$, then according to Eq.~(\ref{con:ZsBoundDecay}),
\begin{equation} \label{con:ZsteadyPr}
\begin{aligned}
\langle \sigma_1^z (\infty)\rangle & = \frac{s\left ( z_0 - \frac{\Gamma_{\mathrm{eff}}'}{s}\right )\left [ \left (s+ \frac{\Gamma_{\mathrm{eff}}'}{2}\right )^2+Y'^2\right ] }{\left [ \left ( s + \frac{\Gamma_{\mathrm{eff}}'}{2}\right )^2 + Y'^2 \right ] \left ( s + \Gamma_{\mathrm{eff}}'\right ) + \bar{\Omega}^2 \left (s + \frac{\Gamma_{\mathrm{eff}}'}{2} \right )}\\
&=\frac{4\gamma_0^2 - \Gamma_{\mathrm{eff}}'^2}{2\bar{\Omega}^2 -4\gamma_0^2 + \Gamma_{\mathrm{eff}}'^2 } > -1,
\end{aligned}
\end{equation}
provided that $\bar{\Omega} >0$. This means that the coherent drive $\bar{\Omega}$ can make the atom be excited rather than let it remain at the ground state. In particular,  when $\bar{\Omega}^2 = 4\gamma_0^2 - \Gamma_{\mathrm{eff}}'^2$,  by Eq. \eqref{con:ZsteadyPr} we have $\langle \sigma_1^z (\infty)\rangle = 1$; in other words, the coherent drive can 
compensate the atom's spontaneous decay and the atom can be perfectly excited.
\end{Proof}

\subsection{Measurement feedback control combined with the coherent feedback}

When the quantum system is measured via homodyne detection at the right end of the semi-infinite waveguide, the measured information is related to fluorescence  emitted by the atom into the waveguide~\cite{OPSB,zhan2022long,buonaiuto2021dynamical,gonzalez2017efficient,olmos2020interaction,strandberg2019numerical,Zoller1}. 
Specifically, when an input filed $\textbf{a}_{\rm in}(t)$ is applied to the quantum system  form the right end of the waveguide, the corresponding output field $\textbf{a}_{\rm out}(t)$ reads~\cite{strandberg2019numerical,Anton2018PRL} 
\begin{equation} \label{eq:InputOutputWaveAtom}
\begin{aligned}
\textbf{a}_{\rm out}(t) = \textbf{a}_{\rm in}(t) +  \left [\gamma_{1L} \sigma_1^-\left (t-\frac{2z_1}{c} \right)-\gamma_{1R} \sigma_1^-\left (t\right)\right].
\end{aligned}
\end{equation}
It is noted that the second component on  the RHS can also be derived by solving the Heisenberg equation of the operator $d_k$, as was done in~\cite{guimond2016chiral}.

Under the similar approximation adopted  in~\cite{zhan2022chirality,guimond2016chiral}, the measurement operator for the atom can be represented as~\cite{InputOutput2013}
\begin{equation} \label{con:M1}
\begin{aligned}
\hat{M}_1&= \left (\gamma_{1R} - \gamma_{1L} e^{i2k_az_1}\right) \sigma_1^-,
\end{aligned}
\end{equation}
where $k_a = \omega_a/c$.  Let $X_M = \hat{M}_1 + \hat{M}_1^{\dag}$. Then the output photon current of the system via homodyne detection reads~\cite{zhan2022chirality,OPSB}
\begin{equation} \label{eq:measurey}
\begin{aligned}
I_c(t) & =  \left\langle X_M \right\rangle   + \frac{1}{\sqrt{\eta}}\frac{dW}{\mathrm{d} t}\\
&= \left\langle \hat{M}_1 + \hat{M}_1^{\dag} \right\rangle   + \frac{1}{\sqrt{\eta }}\frac{dW}{\mathrm{d} t}\\
& = \gamma_{1R} \langle X\rangle  - \gamma_{1L}  \left\langle e^{i2k_az_1} \sigma_1^- + e^{-i2k_az_1} \sigma_1^+\right\rangle  + \frac{1}{\sqrt{\eta }}\frac{dW}{\mathrm{d} t},
\end{aligned}
\end{equation}
where $X$ is given above Eq.~(\ref{con:OneatomMeanOperator}), the bracket $\langle \   \rangle$ represents the expectation value of operators under homodyne measurement, $0<\eta\leq 1$ is homodyne detection efficiency, and
$dW$ is a Wiener increment satisfying $E\left [ dW \right ]= 0$ and $E\left [ dW^2\right ] =\mathrm{d} t$ with $E$ denoting an ensemble average.
\begin{remark}
The output photon current in Eq. \eqref{eq:measurey} can be equivalently represented as
\begin{equation} \label{eq:measurey2}
\begin{aligned}
I_c(t) &  = \left(\gamma_{1R} -  \gamma_{1L} e^{i2k_az_1} \right)\langle \sigma_1^- \rangle +  \left(\gamma_{1R} -  \gamma_{1L} e^{-i2k_az_1} \right)\langle \sigma_1^+ \rangle 
+ \frac{1}{\sqrt{\eta}}\frac{dW}{\mathrm{d} t}.
\end{aligned}
\end{equation}
In particular, when $2k_az_1 = n\pi$,
\begin{equation} \label{eq:measurey3}
\begin{aligned}
I_c(t) &  =\tilde{\gamma}\langle X \rangle 
+ \frac{1}{\sqrt{\eta}}\frac{dW}{\mathrm{d} t},
\end{aligned}
\end{equation}
where $\tilde{\gamma} =\gamma_{1R} - \gamma_{1L} \cos(n\pi)$ is the effective coupling. 
\end{remark}

According to quantum measurement theory, the dynamics of the quantum network under homodyne detection is~\cite{strandberg2019numerical}
\begin{equation} \label{eq:mastereqbetaEta}
\begin{aligned}
\mathrm{d}\rho_c(t)  & = -i [H_{\mathrm{eff}},\rho_c(t)]\mathrm{d}t + \Gamma_{\mathrm{eff}} \mathcal{L}_{\sigma_1^-} \left [ \rho_c(t) \right ] \mathrm{d}t + \mathcal{H}[\tilde{\gamma}\sigma_1^-]\rho_c(t)\sqrt{\eta} dW,
\end{aligned}
\end{equation}
where
$\mathcal{L}_{\sigma_1^-}$ is the Lindblad component as given in Eq.~(\ref{con:masterDissipation}), and for an arbitrary operator $L$, the superoperator
\[
 \mathcal{H}[L]\rho = L \rho(t) + \rho(t) L^{\dag} -  \mathrm{Tr} \left [  \right ( L + L^{\dag}\left )\rho(t) \right ]\rho(t).
\]

In the following, we assume $\eta  = 1$ in the homodyne detection measurement. Then the measurement feedback control equation with an arbitrary system operator $F$  reads~\cite{wiseman2009quantum,wiseman1994quantum,gough2009series,zhang2017quantum}
\begin{equation} \label{eq:measFBmaintext}
\begin{aligned}
&\mathrm{d}\rho_F(t)=  -i [H_{\mathrm{eff}},\rho_F(t)]\mathrm{d}t + \Gamma_{\mathrm{eff}} \mathcal{L}_{\sigma_1^-} \left [ \rho_F(t) \right ]\mathrm{d}t  + \mathcal{H}[\tilde{\gamma}\sigma_1^-]\rho_F(t) dW(t)\\
&- i   g_f \left [F, \rho_F(t)\right ] d W(t)  - i g_f \tilde{\gamma}\left [F, \sigma_1^-\rho_F(t) + \rho_F(t)\sigma_1^+\right ] \mathrm{d} t -\frac{1}{2}g_f^2 \left [ F , \left [F, \rho_F(t)\right ] \right ]\mathrm{d} t,
\end{aligned}
\end{equation}
where $g_f$ represents the feedback strength.

Denote $\xi(t) = dW/\mathrm{d} t$. When $H_{\mathrm{eff}} = -\left ( Y + i\gamma_0 \right ) \sigma_1^+ \sigma_1^- + \left(\bar{\Omega}_1/2\right)X$ as in Eq. \eqref{con:masterDissipation} with $\gamma_0 \neq 0$, 
$F = X$~\cite{OPSB},  by the stochastic master equation  \eqref{eq:measFBmaintext}, the  evolution of the  atomic operators reads
\begin{footnotesize}
\begin{subequations} \label{con:MFOneAtomOperator}
\begin{align}
&\frac{\mathrm{d}}{\mathrm{d}t} \langle X \rangle = i \left (Y+i\gamma_0\right) \left ( \langle \sigma_1^-\rangle -\langle \sigma_1^+\rangle \right )-\frac{\Gamma_{\mathrm{eff}}}{2}\left ( \langle \sigma_1^-\rangle +\langle \sigma_1^+\rangle \right ) + \tilde{\gamma} \left (1+ \langle \sigma_1^z \rangle - \langle X \rangle^2\right )\xi(t),   \label{FBOneAtomMean1}\\
&\frac{\mathrm{d}}{\mathrm{d}t} \langle \sigma_1^+ \rangle = -i \left (Y+i\gamma_0 \right )\langle \sigma_1^+\rangle -\frac{\Gamma_{\mathrm{eff}}}{2}\langle \sigma_1^+\rangle - \frac{i\bar{\Omega}}{2}\langle \sigma_1^z \rangle  +ig_f\tilde{\gamma} \langle X\rangle  + g_f^2 \left (\langle \sigma_1^-\rangle - \langle \sigma_1^+\rangle \right )\notag\\
&~~~~~~~~~~~~~+ \tilde{\gamma} \left ( \frac{1+\langle \sigma_1^z\rangle}{2} - \langle X\rangle \langle \sigma_1^+\rangle \right )\xi(t)-ig_f \langle \sigma_1^z \rangle \xi(t) ,   \label{FBOneAtomMean2}\\
&\frac{\mathrm{d}}{\mathrm{d}t} \langle \sigma_1^- \rangle = i \left (Y+i\gamma_0 \right )\langle \sigma_1^-\rangle -\frac{\Gamma_{\mathrm{eff}}}{2}\langle \sigma_1^-\rangle + \frac{i\bar{\Omega}}{2}\langle \sigma_1^z \rangle  -ig_f\tilde{\gamma} \langle X\rangle + g_f^2\left (\langle \sigma_1^+\rangle - \langle \sigma_1^-\rangle\right) \notag\\  
&~~~~~~~~~~~~~ + \tilde{\gamma} \left ( \frac{1+\langle \sigma_1^z\rangle}{2} - \langle X\rangle \langle \sigma_1^-\rangle \right )\xi(t) + ig_f \langle \sigma_1^z \rangle \xi(t), \label{FBOneAtomMean3}\\
&\frac{\mathrm{d}}{\mathrm{d}t} \langle \sigma_1^z \rangle = -i\bar{\Omega} \left (\langle \sigma_1^+ \rangle -\langle \sigma_1^- \rangle \right) -\Gamma_{\mathrm{eff}} \left (\langle \sigma_1^z \rangle +1 \right) - 2g_f^2 \langle \sigma_1^z \rangle \notag\\ 
&~~~~~~~~~~~~~ - \tilde{\gamma}\langle X\rangle \left ( 1+ \langle \sigma_1^z \rangle \right ) \xi(t) -2ig_f \left ( \langle \sigma_1^+ \rangle -\langle \sigma_1^- \rangle \right )\xi(t) .    \label{FBOneAtomMean4}
\end{align}
\end{subequations}
\end{footnotesize}%

Averaging over $\xi(t) = 0$ and applying the Laplace transform to $\langle \sigma_1^z\rangle$,  similar as in Eq.~(\ref{con:XLaplace3}),  yields
\begin{equation} \label{con:ZsMeasurementFB}
\begin{aligned}
\mathcal{Z}(s)=\frac{z_0 - \Gamma_{\mathrm{eff}}/s}{\left ( s + \Gamma_{\mathrm{eff}} +2g_f^2\right) + \frac{2 g_f^2 \bar{\Omega}^2 +\bar{\Omega}^2 \left(s + \Gamma_{\mathrm{eff}}/2 \right ) }{ \left (s+ \Gamma_{\mathrm{eff}}/2 + g_f^2  \right )^2 - \mathbf{F}_z -g_f^4}},
\end{aligned}
\end{equation}
where $\mathbf{F}_z =\left(\gamma_0 -iY\right)\left(\gamma_0 -iY+2ig_f \tilde{\gamma}\right) $.

It can be seen that the atomic state is determined by the decaying process to the waveguide and environment, the coherent drive $ \bar{\Omega}$ and the measurement feedback control. Then we have the following theorem.

\begin{thm} \label{Theorem6MF}
The measurement feedback control cannot drive the atom to be fully excited; instead  it can shorten its final distance to the state for which $\langle \sigma_1^z\rangle = 0 $, and the distance is determined by the feedback strength $g_f^2$ if $Y, \gamma_0 \ll g_f \tilde{\gamma}$.
\end{thm}
\begin{Proof}
By \textbf{Theorem~\ref{QiTAC}} and \textbf{Remark~\ref{QiRemark}}, the atom cannot be repeatedly excited.   
We consider the steady value 
\begin{equation} \label{con:ZSteadyPM}
\begin{aligned}
\langle \sigma_1^z (\infty)\rangle & = \lim_{s\rightarrow 0} s \mathcal{Z}(s) \\
&=\frac{ - \Gamma_{\mathrm{eff}}}{\left (  \Gamma_{\mathrm{eff}} +2g_f^2\right) + \frac{\bar{\Omega}^2 \left(2 g_f^2 + \frac{\Gamma_{\mathrm{eff}}}{2} \right ) }{\frac{\Gamma_{\mathrm{eff}} }{2}  \left ( 2g_f^2 + \frac{\Gamma_{\mathrm{eff}} }{2}   \right ) - \mathbf{F}_z }}<0.
\end{aligned}
\end{equation}
When $Y, \gamma_0 \ll g_f \tilde{\gamma}$, $g_f$ can influence the steady value of $\langle\sigma_1^z\rangle$ mainly by the first component of the 
denominator of Eq.~(\ref{con:ZSteadyPM}) and $g_f$ can shorten the distance between $\langle \sigma_1^z (\infty)\rangle$ and zero.
\end{Proof}

\begin{remark} \label{NoOmega}
For a special case that $\bar{\Omega} = 0$, which means there is no coherent drive, then $\langle \sigma_1^z (\infty)\rangle  =-\frac{\Gamma_{\mathrm{eff}}}{\Gamma_{\mathrm{eff}} + 2g_f^2 }$.
\end{remark}

The following theorem tells how the homodyne detection noise can influence the amplitudes of quantum state around its final steady state.

\begin{thm}\label{Cohestability}
If $g_f \gg \gamma_{1R} \geq \gamma_{1L}$, then the chiral couplings between the atom and the waveguide enhance stochastic fluctuations  in the amplitude of the atom.
\end{thm}
\begin{Proof}
When $\xi(t)\neq 0$, we denote $\langle \sigma_1^+ \rangle =  \langle \bar{\sigma}_1^+  \rangle + \Delta\sigma_1^+ $, $\langle \sigma_1^- \rangle =  \langle \bar{\sigma}_1^-  \rangle  + \Delta\sigma_1^- $, $\langle \sigma_1^z \rangle = \langle \bar{\sigma}_1^z  \rangle + \Delta\sigma_1^z $, where $ \langle \bar{\sigma}_1^+  \rangle $, $ \langle \bar{\sigma}_1^-  \rangle $ and $ \langle \bar{\sigma}_1^z  \rangle $ represent the mean value of the operators in Eq.~(\ref{con:MFOneAtomOperator}) when $\bar{\Omega} = 0$ and $\xi(t) = 0$. Thus $\Delta\sigma_1^+$, $\Delta\sigma_1^z$ and $\Delta\sigma_1^z$ are the stochastic values induced by $\xi(t) \neq 0$. Similar as Eq.~(\ref{con:ZSteadyPM}), we can derive that $ \langle \bar{\sigma}_1^+  \rangle  =  \langle \bar{\sigma}_1^-  \rangle  =  \langle \bar{\sigma}_1^z  \rangle  = 0$. When $g_f \gg \Gamma_{\mathrm{eff}}$, the dominant stochastic component in Eq.~(\ref{FBOneAtomMean4}) is proportional to  
\begin{equation} \label{con:domainStochastic}
\begin{aligned}
\langle \sigma_1^+ \rangle -\langle \sigma_1^- \rangle  &= \Delta\sigma_1^+ - \Delta\sigma_1^- =-2ig_f  \langle \sigma_1^z \rangle  dW \\
&=-2ig_f \left ( \langle \bar{\sigma}_1^z  \rangle  + \Delta\sigma_1^z \right )dW \\
&= -2ig_f \langle \bar{\sigma}_1^z  \rangle   dW \\
&= 2ig_f\frac{\Gamma_{\mathrm{eff}}}{\Gamma_{\mathrm{eff}} + 2g_f^2 } dW,
\end{aligned}
\end{equation}
where $\Delta\sigma_1^zdW$ is omitted because $\mathrm{d}W^{2+n} = 0$ for $n > 0$~\cite{gardiner1985handbook}. According to Eq.~(\ref{con:ZSteadyPM}) with $\bar{\Omega}=0$, the final steady value is determined by $V = g_f^2/\Gamma_{\mathrm{eff}}$. For given $V$ with $g_f \gg \Gamma_{\mathrm{eff}}$, we have 
\begin{equation} \label{con:domainStochasticCal2}
\begin{aligned}
\langle \sigma_1^+ \rangle -\langle \sigma_1^- \rangle & = 2ig_f\frac{\Gamma_{\mathrm{eff}}}{\Gamma_{\mathrm{eff}} + 2g_f^2 } dW =2i\frac{\sqrt{V\Gamma_{\mathrm{eff}}}}{1+2V}dW.
\end{aligned}
\end{equation}
This means that the settings $\gamma_{1L} \neq \gamma_{1R}$ with large $\Gamma_{\mathrm{eff}}$ can induce large stochastic
fluctuations of the quantum state around its steady value.
\end{Proof}
\begin{thm}\label{transition}
When $\omega_a z_1 = n\pi$ where $n=1,2,\cdots$, chiral couplings between the atom and waveguide can induce faster evolution of average atomic transients when $\bar{\Omega}=0$.
\end{thm}

\begin{Proof}
Consider the dynamics in Eq.~(\ref{FBOneAtomMean4}) after averaging $\xi(t) $. 
When $\omega_a z_1 = n\pi$, $\Gamma_{\mathrm{eff}}$ is real and  positive.  By Eq.~(\ref{FBOneAtomMean4}),
\begin{equation} \label{con:ZsTheore8}
\begin{aligned}
\langle \sigma_1^z (\infty)\rangle & \approx\frac{ - \Gamma_{\mathrm{eff}}}{  \Gamma_{\mathrm{eff}} +2g_f^2 }
=- \frac{1}{1+2V },
\end{aligned}
\end{equation}
where $g_f = \sqrt{\Gamma_{\mathrm{eff}} V}$. By  Eq. \eqref{con:ZsTheore8} the atom converges to the same steady state for a given $V$. 
Let $\gamma_{1R}+\gamma_{1L}$ be a constant, $\Gamma_{\mathrm{eff}}$ will be larger when $\gamma_{1R} \neq \gamma_{1L}$.
Then according to  Eq.~(\ref{FBOneAtomMean4}),  $\langle\sigma_1^z (t) \rangle$ approaches its steady value with a faster convergence rate. 
\end{Proof}

\begin{figure}[h]
\centerline{\includegraphics[width=1\columnwidth]{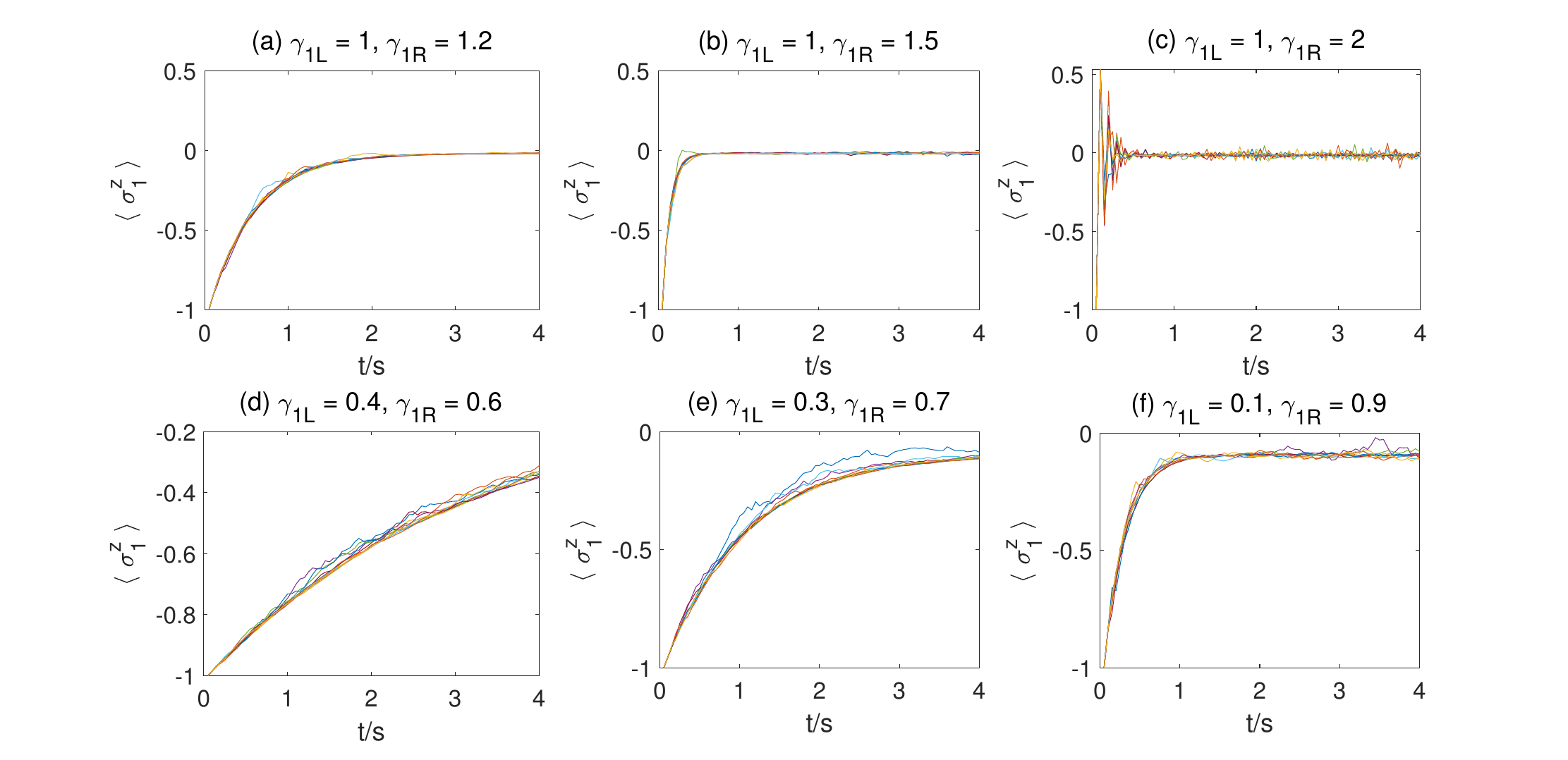}}
\caption{Measurement feedback control influenced by coherent feedback designs.}
	\label{fig:Stability}
\end{figure}

In the following, we illustrate  \textbf{Theorem~\ref{Cohestability}} and \textbf{Theorem~\ref{transition}}. 
Let $\omega_a = 50$, $\gamma_0 = 0$, $\bar{\Omega} = 0$, $\Gamma = 0.01$ and $\mathrm{d} t = 0.05s$. Assume that initially the atom is at the ground state. 
In Fig.~\ref{fig:Stability}(a-c),  $g_f^2/\Gamma_{\mathrm{eff}} = 30$. It is clear that   the atom finally converges to the same steady value. When $z_1 = \pi/\omega_a$, the chiral couplings between the atom and waveguide can induce large $\Gamma_{\mathrm{eff}}$, ($\Gamma_{\mathrm{eff}} = 0.03, 0.135, 0.51$ in (a-c), respectively),  which lead to large stochastic fluctuation. This agrees with the conclusion in \textbf{Theorem~\ref{Cohestability}}. In Fig.~\ref{fig:Stability}(d-f), the parameters are the same as (a-c) except the coupling strengths and $g_f^2/\Gamma_{\mathrm{eff}} = 5$. According to the chiral coupling parameters, we have  $\Gamma_{\mathrm{eff}} = 0.03$ in (d), $\Gamma_{\mathrm{eff}} = 0.09$  in (e)  and  $\Gamma_{\mathrm{eff}} = 0.33$  in (f). The simulations clearly show that $\langle\sigma_1^z\rangle$ converges faster with larger $\Gamma_{\mathrm{eff}}$, which  agrees with the conclusion in \textbf{Theorem~\ref{transition}}.

\section{Conclusions}
\label{sec:conclusions}

In this paper, we have studied the feedback control of a quantum system where two-level atoms are coupled to a semi-infinite waveguide. When initially there are no more than two atoms that are excited and the waveguide is empty, the quantum state  evolution can be interpreted as the evolution of a random graph. By adjusting the location of the atoms and the chiral coupling strengths between the atoms and the waveguide, the number of photons and the consensus of the atomic dynamics  can be controlled.
Besides, the spontaneous emission and dissipation of the atoms induced by the environment can influence the steady state, and this process can be compensated by external coherent drives and measurement feedback controls. Moreover, the coherent feedback can be tuned by designing the chiral couplings and the location of the atoms, and this, together with  measurement feedback control,   influences the oscillation of atomic states.  

\appendix
\section{Derivation of the delay dependent control equation with one and two excitations based on the Schr\"{o}dinger equation} \label{Sec:ApendixDelayOnexcitation}
In this section, we derive the delay dependent control equation in the main text according to the number of possible excitations emitted by the atoms in the waveguide, Specifically,  Section~\ref{sec:onephotonDerivation} provides the details of the derivation of Eq.~(\ref{con:WavNatomDelay}) with one excitation and section~\ref{Sec:ApendixDelay} is devoted to the derivation of Eq.~(\ref{con:PopuquationDelay}) with two excitations.

\subsection{Derivation of Eq.~(\ref{con:WavNatomDelay})} \label{sec:onephotonDerivation}
The integration of Eq.~(\ref{ckt}) into Eq.~(\ref{cjt}) gives
\begin{equation} \label{con:StateAppendix}
\begin{aligned}
&~~~~\dot{c}_j(t) = -i\int g_{kjt}^*(k,t,z_j) \tilde{c}(t,k) \mathrm{d}k\\
&=\int \left \{ -\gamma_{jR} e^{-i\left [\left (\omega_k-\omega_a\right )t - \omega_k z_j/c\right ]} +\gamma_{jL}e^{-i\left [\left (\omega_k-\omega_a\right )t + \omega_k z_j/c\right ]}\right \}\\
 &~~~~\sum_{p=1}^N  \int_0^t c_p(u) \left \{ \gamma_{pR} e^{i\left [\left (\omega_k-\omega_a\right )u - \omega_k z_p/c\right ]} - \gamma_{pL}e^{i\left [\left (\omega_k-\omega_a\right )u + \omega_k z_p/c\right ]}\right \}\mathrm{d}u \mathrm{d}k\\
&=- \frac{\gamma_{jR}^2 + \gamma_{jL}^2}{2}c_j(t)-\gamma_{jR} \sum_{p=1}^{j-1}  \gamma_{pR}  c_p\left(t-\frac{z_j-z_p}{c}\right)e^{i\omega_a\frac{z_j-z_p}{c}}\\
&~~~~+\gamma_{jR} \sum_{p=1}^N  \gamma_{pL} c_p\left(t-\frac{z_p+z_j}{c}\right)  e^{i\omega_a \frac{z_p+z_j}{c}} \\
&~~~~-\gamma_{jL} \sum_{p=j+1}^N \gamma_{pL} c_p\left(t-\frac{z_p-z_j}{c}\right)  e^{i\omega_a\frac{z_p-z_j}{c}}.
\end{aligned}
\end{equation}
Next, we take one of the integrals in the second and third lines of Eq.~(\ref{con:StateAppendix}) as an example. We have
\begin{equation} \label{con:StateExample}
\begin{aligned}
&~~~~\gamma_{jR}\int_0^t c_p(u) \int \gamma_{pR} e^{i\left [\left (\omega_k-\omega_a\right )u - \omega_k z_p/c\right ]}  e^{-i\left [\left (\omega_k-\omega_a\right)t - \omega_k z_j/c\right]} \mathrm{d}k \mathrm{d}u\\
&=\gamma_{jR}  \int_0^t c_p(u) \gamma_{pR} e^{i\omega_a(t-u)} \delta\left (u-t-z_p/c + z_j/c\right )\mathrm{d}u\\
&=\gamma_{jR} \gamma_{pR}  c_p\left (t-\frac{z_j-z_p}{c}\right )e^{i\omega_a\frac{z_j-z_p}{c}}.
\end{aligned}
\end{equation}
The other components in Eq.~(\ref{con:WavNatomDelay}) can be similarly derived.

\subsection{Derivation of Eq.~(\ref{con:WavNatomDelay})} \label{Sec:ApendixDelay}

Plugging Eq.~(\ref{NAmodel3}) into Eq.~(\ref{NAmodel2}), then the second component of the RHS of Eq.~(\ref{NAmodel2}) reads
\begin{footnotesize}
\begin{equation} \label{con:Eq2Component2}
\begin{aligned}
&~~~~i  \int c_{*}(k,k_2,t)g_{kjt}^*(k_2,t,z_j) \mathrm{d}k_2\\
&= \int \int_0^t\sum_{j'} \left [c_{j'}^{\bullet}(u,k)  g_{kj't}(k_2,u,z_j) +c_{j'}^{\bullet}(u,k_2)  g_{kj't}(k,u,z_{j'}) \right ]\mathrm{d}u g_{kjt}^*(k_2,t,z_j) \mathrm{d}k_2\\
&= \sum_{j'} \int \int_0^t \left [c_{j'}^{\bullet}(u,k)  g_{kj't}(k_2,u,z_{j'}) +c_{j'}^{\bullet}(u,k_2)  g_{kj't}(k,u,z_{j'}) \right ] g_{kjt}^*(k_2,t,z_j) \mathrm{d}u\mathrm{d}k_2\\
&= \sum_{j'} \int_0^t \int c_{j'}^{\bullet}(u,k) \left (\gamma_{j'R} e^{-ik_2 z_{j'}}-\gamma_{j'L} e^{ik_2 z_{j'}}\right )\\
&~~~~\left(\gamma_{jR} e^{ik_2 z_{j}}-\gamma_{jL} e^{-ik_2 z_{j}}\right ) e^{ik_2c(u-t)}\mathrm{d}k_2e^{i\omega_a(t-u)}\mathrm{d}u \\
&= \sum_{j'} \int_0^t  c_{j'}^{\bullet}(u,k) \left [ \gamma_{j'R} \gamma_{jR} \delta\left(u-t+\frac{z_j-z_{j'}}{c}\right) -\gamma_{j'R} \gamma_{jL} \delta\left(u-t-\frac{z_j+z_{j'}}{c}\right) \right.\\
&~~~~ \left.- \gamma_{j'L}\gamma_{jR} \delta\left(u-t+\frac{z_j+z_{j'}}{c}\right) 
 + \gamma_{j'L}\gamma_{jL} \delta\left(u-t+\frac{z_{j'}-z_j}{c}\right) \right]e^{i\omega_a(t-u)}\mathrm{d}u\\
&=\frac{\gamma_{jR}^2 + \gamma_{jl}^2}{2}c_{j}^{\bullet}(t,k) - \sum_{j'}\gamma_{j'R} \gamma_{jL}c_{j}^{\bullet}\left(t-\frac{z_j+z_{j'}}{c},k\right) e^{i\omega_a\frac{z_j+z_{j'}}{c}}\\
&~~~~+ \sum_{j'<j}\gamma_{j'R} \gamma_{jR} c_j^{\bullet}\left(t-\frac{z_j-z_{j'}}{c}\right) e^{i\omega_a\frac{z_j-z_{j'}}{c}} \\
&~~~~+\sum_{j'>j}\gamma_{j'L}\gamma_{jL} c_{j}^{\bullet}\left(t-\frac{z_{j'}-z_j}{c},k\right)e^{i\omega_a\frac{z_{j'}-z_j}{c}},
\end{aligned}
\end{equation}
\end{footnotesize}%
which is the delay-dependent evolution format in Eq.~(\ref{NAmodel2Delay}).

Then take the first component of RHS of Eq.~(\ref{NAmodel1}) as an example,
\begin{footnotesize}
\begin{equation} \label{con:Eq1Component1}
\begin{aligned}
&~~~~\int c_{j}^{\bullet}(t,k)g_{klt}^*(k,t,z_l)\mathrm{d}k\\
&=\sum_{l'} \int_0^t \int -i c_{jl'}(u)  g_{kl't}(k,u,z_{l'}) g_{klt}^*(k,t,z_l)\mathrm{d}k\mathrm{d}u\\
&= -i \sum_{l'} \int_0^t \int c_{jl'}(u) \left [ \gamma_{l'R} \gamma_{lR}e^{i\omega_k\left (u - t + \frac{z_l-z_{l'}}{c}\right)} - \gamma_{l'R} \gamma_{lL}e^{i\omega_k\left (u - t - \frac{z_l+z_{l'}}{c}\right)} \right. \\
&~~~~\left.-  \gamma_{l'L} \gamma_{lR} e^{i\omega_k\left (u - t + \frac{z_l+z_{l'}}{c}\right)} + \gamma_{l'L} \gamma_{lL} e^{i\omega_k\left (u - t + \frac{z_{l'}-z_l}{c}\right)}  \right]e^{i\omega_a(t-u)}\mathrm{d}k\mathrm{d}u \\
&= -i\sum_{l'=1}^{l-1} \int_0^t  \gamma_{l'R} \gamma_{lR}c_{jl'}(u) \delta\left (u - t + \frac{z_l-z_{l'}}{c}\right)e^{i\omega_a(t-u)}\mathrm{d}u \\
&~~~~ + i \sum_{l'=1}^{l-1} \int_0^t c_{jl'}(u)\gamma_{l'L} \gamma_{lR} \delta\left (u - t + \frac{z_l+z_{l'}}{c}\right) e^{i\omega_a(t-u)}\mathrm{d}u \\
&~~~~-i\sum_{l'=l+1}^{N} \int_0^t \gamma_{l'L} \gamma_{lL} c_{jl'}(u) \delta\left (u - t + \frac{z_{l'} -z_l}{c}\right)e^{i\omega_a(t-u)}\mathrm{d}u -i\frac{\gamma_{lR}^2 +\gamma_{lL}^2}{2}c_{jl}(t) \\
&= -i\sum_{l'=1,l'\neq j}^{l-1}  \gamma_{l'R} \gamma_{lR}c_{jl'}\left (t - \frac{z_l-z_{l'}}{c}\right)e^{i\omega_a\frac{z_l-z_{l'}}{c}} \\
&~~~~ + i \sum_{l'=1,l'\neq j}^{N}  \gamma_{l'L} \gamma_{lR} c_{jl'}\left (t - \frac{z_l+z_{l'}}{c}\right) e^{i\omega_a\frac{z_l+z_{l'}}{c}} \\
&~~~~-i\sum_{l'=l+1,l'\neq j}^{N} \gamma_{l'L} \gamma_{lL} c_{jl'}\left ( t - \frac{z_{l'} -z_l}{c}\right)e^{i\omega_a\frac{z_{l'} -z_l}{c}}\mathrm{d}u -i\frac{\gamma_{lR}^2 +\gamma_{lL}^2}{2}c_{jl}(t),
\end{aligned}
\end{equation}
\end{footnotesize}%
which is half of the components in Eq.~(\ref{NAmodel1Delay}) and the other half can be similarly derived.

\bibliography{ref}

\end{document}